\documentclass[prc,aps,preprint]{revtex4}

\RequirePackage{lineno}
\usepackage{graphicx}% Include figure files
\usepackage{dcolumn}% Align table columns on decimal point
\usepackage{bm}% bold math
\usepackage{slashed}
\usepackage{ulem}
\usepackage{url}
\usepackage{amsmath}
\usepackage{amssymb}
\usepackage{color}

\RequirePackage{graphicx}
\RequirePackage{fix-cm}
\RequirePackage{latexsym}
\RequirePackage[colorlinks,citecolor=blue,urlcolor=blue,linkcolor=blue]{hyperref}

\begin{document}

\title{Residual Mean Field Model of  Valence Quarks in the Nucleon}
\author{ Christopher Leon and Misak Sargsian}
\affiliation{ Department of Physics, Florida International University, Miami, Florida 33199, USA}
\date{\today}

%\title{Residual Mean Field Model of  Valence Quarks in the Nucleon}

%\author{Christopher Leon\thanksref{e1,addr1}
%        \and
%        Misak Sargsian\thanksref{addr1}
%}

%\thankstext{e1}{e-mail: cleon082@fiu.edu}

%\institute{ Department of Physics, Florida International University, Miami, Florida 33199, USA \label{addr1}
%}

%\maketitle

\begin{abstract}
We develop a non-perturbative model for valence parton distribution functions (PDFs)  based on the mean field interactions 
of valence quarks  in  the nucleonic interior. 
The main motivation for the model is to obtain a mean field description of the valence quarks as a baseline to study the 
short range quark-quark interactions that generate  the high $x$ tail of PDFs. 
The model is based on the separation of the valence three-quark cluster and  residual system in the nucleon. 
Then the  nucleon structure function is calculated  within the effective light-front diagrammatic approach  introducing 
nonperturbative  light-front valence quark and residual wave functions.  
Within  the model a new relation is obtained  between the position, $x_p$,  of the peak of 
$xq_V(x)$ distribution of the valence quark and the effective mass of the residual system, $m_R$,  in the form: 
$x_{p} \approx {1\over 4} (1-{m_R\over m_N})$ at starting $Q^2$.
This relation explains the difference in the peak positions for d- and u- quarks through the expected difference of 
residual masses for valence d-  and u- quark distributions.
The parameters of the model are fixed   by fitting the calculated valence quark distributions to the phenomenological PDFs.
This allowed us to estimate the  overall  mean field contribution in baryonic and momentum sum rules for valence d- and u- quarks.
Finally,  the  evaluated   parameters of  the non-perturbative wave functions of valence 3q-cluster  and residual system can be used in calculation of other quantities such as nucleon form factors,  generalized partonic and transverse momentum distributions.
\end{abstract}
\maketitle

\section{Introduction}
Understanding the dynamics of the valence quarks in the nucleon is one of the most interesting aspects of quantum chromodynamics~(QCD).  
The problem is not only how the valence PDFs can be calculated  from the first principles in QCD, but also 
how in  general  the three relativistic Fermions become bound. 
The special feature of valence quarks is that they define the baryonic  quantum number of the  nucleon 
and one expects them to be a bridge between baryonic spectroscopy and the QCD structure of the nucleon.
Since baryonic spectroscopy adheres reasonably well to the SU(6) spin-flavor symmetry, one expects such a symmetry to be found also in the 
valence quark distributions in the nucleon  and the problem is to 
understand how and in which parts of the valence quark momentum distribution this symmetry is broken.

Historically, the first attempts to model valence quark structure of nucleon were based on the non-relativistic picture of constituent quarks in 
which SU(6) symmetry was preserved  for the  constituents carrying approximately one third of the nucleon mass\cite{Isgur:1979be,Close:1979bt}.  
These models were  successful in describing the  phenomenologies of baryonic spectroscopy. 
However, experimental extraction of the valence PDFs  indicates  that SU(6) is  apparently  violated, especially at large Bjorken $x$.  
Thus, one of the unresolved issues  is how to reconcile 
the apparent success of SU(6) in baryonic spectroscopy and its breaking down in partonic distributions.

Another unique property of valence quarks is that their distribution weighted by Bjorken $x$ has a well defined  peak, 
even if the shape of the PDF is  not an observable. 
The peaking of the momentum distribution for the bound system  implies the importance of mean-field dynamics for the individual 
valence fermions  in the nucleon, while the position of the peak is sensitive to the dynamical aspects of interaction 
in the mean field

The history of the modeling of partonic distributions is very rich, ranging from non-relativistic\cite{Isgur:1979be} and  relativistic\cite{Brodsky:1981jv} 
constituent quark models,  bag models\cite{Jaffe:1974nj,Chodos:1974pn,Miller:1979kg}, models combining partonic and pionic cloud picture of 
the nucleon\cite{Thomas:1981vc,Schreiber:1991qx,Miller:2002ig} as well as models based on the  di-quark picture of the nucleon\cite{Close:1988br,Anselmino:1992vg,Roberts:1994dr,Cloet:2005pp,Cloet:2013jya} following from the spin-dependent part of the one-gluon exchange in 
the non-relativistic approximation\cite{DeRujula:1975qlm}.
All these models are non-perturtbative in their nature and their predictions varied widely for the general characteristics of the valence quark distribution such as
position of the peak and relative strength of the $u$- and $d$-quark distributions. While calculations based on lattice QCD reproduce the general characteristics of valence PDFs, their complexity does not always allow an understanding of the dominant mechanisms of interactions.

In this and the following papers we develop a new model for valence quark distributions of the nucleon based on the multi-quark correlation picture. 
The validity of such a model is based on the fact that even though the number of the quarks is not conserved in the nucleon, the number of 
valence quarks is effectively conserved and therefore it is possible to describe them in the framework used for the description of a bound system of  finite  number of fermions. 
This approach is similar to  highly successful multi-nucleon correlation model of calculation of momentum distribution of 
nucleons in the  nuclei\cite{Frankfurt:2008zv,Egiyan:2005hs,Frankfurt:1988nt,Arrington:2011xs,Fomin:2017ydn,Artiles:2016akj} as well as the 
calculation of the momentum distribution in ultra-cold 
atomic Fermi gas with contact interactions (see e.g. \cite{Tan2008,Hen:2014lia}).
In our approach we consider three distinct interaction dynamics of valence quarks which are: mean field 
three valence quark (3q) cluster, two-quark and three-quark short range interactions, all of which are expected to dominate at different  momentum fraction ranges of the valence quarks. 
We demonstrate that such a framework in the 
description of the valence quark dynamics brings a  new insight and methodology in analyzing the partonic 
distributions in the nucleon.

Why modeling? The advantage of the proposed framework is  that it creates a new ground for making predictions for different QCD processes 
involving nucleons, since  in this case one can make unique predictions based on whether the process under study is dominated by the 
interaction of quarks in the mean field or in two-/three- quark correlations.   As experience from  nuclear physics shows,  eventually the ab-initio calculations\cite{Schiavilla:2006xx,Wiringa:2013ala} (lattice calculation in the case of QCD) will reproduce the phenomena observed based 
on the model description of the processes\cite{Piasetzky:2006ai,Sargsian:2012sm,Hen:2014nza}. 
However, ab-initio calculations,  because of their 
general approach, are not always well positioned for making experimentally verifiable specific predictions.

The article is organized as follows: in Sec.\ref{phenom} we  first present  the justification for the mean field, two- and three- quark  short range 
interaction picture of partonic distributions.  We then discuss the phenomenology of partonic distributions, which are almost universal for 
the all  recent  PDFs extracted from the analysis of different high energy scattering data of electro-production and weak interaction 
processes (e.g. deep inelastic scattering and Drell-Yan).   Our focus is on the dynamical  characteristics of the $x$ weighted valence PDFs, such as the position and height of the peaks  as well as the ratio  of d- to  u- quark distributions. 
We examine how the the position and the height of the peak changes due to QCD 
evolution while the difference between  the peak positions for u- and d- quark distribution is largely $Q^2$ independent. 
Another observation is  the approximate validity of SU(6) symmetry in the  region close to the peak of the partonic distributions. 
The latter observation is used as a justification for the assumption of approximate SU(6) symmetry for mean-field valence quarks.
In Sec.\ref{model}  the mean field model of valence quark distributions is presented.  
Based on this model we calculate the $x q_V(x,Q^2)$ in leading order (LO), which is then  fitted   to the phenomenological distributions (in Sec.\ref{estimates}) to evaluate the parameters defining the non-perturbative wave functions.
The latter allowed us to ascertain the expected strength of the  two- and three- quark correlation effects  in the normalization and momentum sum rule of  valence d- and u- quark distributions. The evaluated parameters of 
the wave functions can also be used for estimation of mean field contributions of the different QCD processes sensitive to 
the valence quark dynamics at moderate Bjorken x $ \approx 0.2$.
Sec.\ref{outlook} summarizes our results,  discusses  the predictions  that follow  from the model and  the possibility of their
experimental verification. We also  discuss the limitations of the model  and outline the second part of the work 
in which the mechanism of quark-quark short range interaction
is included to calculate the high momentum component of valence quark distributions.
In Appendix A we summarize the Light-Front effective diagrammatic rules on which the calculation in the paper is based. 
In Appendix B we present the mathematical details of the derivation related to the integration in the transverse momentum space.

\section{The model of  valence quark distributions and phenomenology of PDFs}
\label{phenom}
The uniqueness of the valence quarks as carriers of the baryonic quantum number of nucleons is that even if the number of quarks 
in the bound nucleon is not conserved their ``effective" number is conserved.  This situation allows us to introduce  a concept of 
potential energy  and  apply a rather well known 
theoretical  framework for  the description of a bound system consisting of a finite number of fermions.
In this framework the bulk of the momentum distribution is defined by the mean  field  dynamics while,
if interaction is short range, the fermion-fermion short-range correlations are responsible for the high momentum  
part of the same distribution. As it was mentioned  in the introduction,  this approach has been successfully 
applied in nuclear physics and  physics of ultra-cold atomic Fermi gases.

\subsection{Mean Field and Quark Correlation Model of Valence Quark Distributions in the Nucleon}
\label{totalmodel}

Any  finite size, bound system with a fixed  number of  fermions with nonzero interaction length will   
exhibit a mean field interaction of its constituents \cite{Migdal:1967ab}.  
Hereafter,  we call a  mean field an approximation  in which the interaction of a given constituent 
with other  constituents in the system  is characterized by a single  effective potential representing sum of 
all possible non-perturbative interactions between constituents. (Such a condition is similar to one in which 
Hartree-Fock approximation is justified.)
In quantum mechanics if the binding of a constituent is due to the minimum of the mean field potential, 
the latter can be approximated by a harmonic oscillator potential near the vicinity of the minimum. 
Thus the   momentum distribution of  the   constituent in 
such a  mean field can be described in  exponential form with the coefficient of the exponent characterizing the 
 size of the system (or orbit). 
 The same mean field dynamics define also 
 the characteristic  average  momentum  carried by the constituents which  can be   related to the position of the 
 peak of the  momentum weighted distributions. 
For the system in which interaction strength between constituents is not negligible  at distances much 
smaller than the bulk size of the system (like the contact interaction, see e.g. \cite{Hen:2014lia}), the 
high momentum part of the momentum distributions is a  result of  the short range pair-wise interactions between constituents, 
often referred to as short-range correlations.
 
 In the present work we apply such a framework for the calculation of  the light-front momentum distribution of valence quarks in the 
 nucleon\cite{Leon:2020cev}. The approach is based on the argument, that the ``effective" number of valence quarks is conserved and the 
strength of the quark-quark interactions is sufficiently large  in the volume they occupy.  It is  assumed that the volume 
the valence quarks occupy  is smaller than the actual size of the nucleon.  Thus we arrive at the picture of  a non-perturbative 
valence quark cluster  embedded in the nucleon. We assume the following interaction dynamics for valence quarks in the nucleon:
the non-perturbative interaction among the three valence quarks provides the confinement 
(referred hereafter as mean field interaction), the short range 
perturbative quark-quark interactions through hard gluon exchanges generate the high momentum  (high Bjorken x) tail of the valence quark 
distribution  (hereafter, referred to  as quark-quark short range correlations)  and the interaction of three-valence quark cluster  
with the residual nucleon system, R,   influences  the final momentum distribution of  valence quarks in the nucleon.

\begin{figure*}
\includegraphics[width=1.0\textwidth]{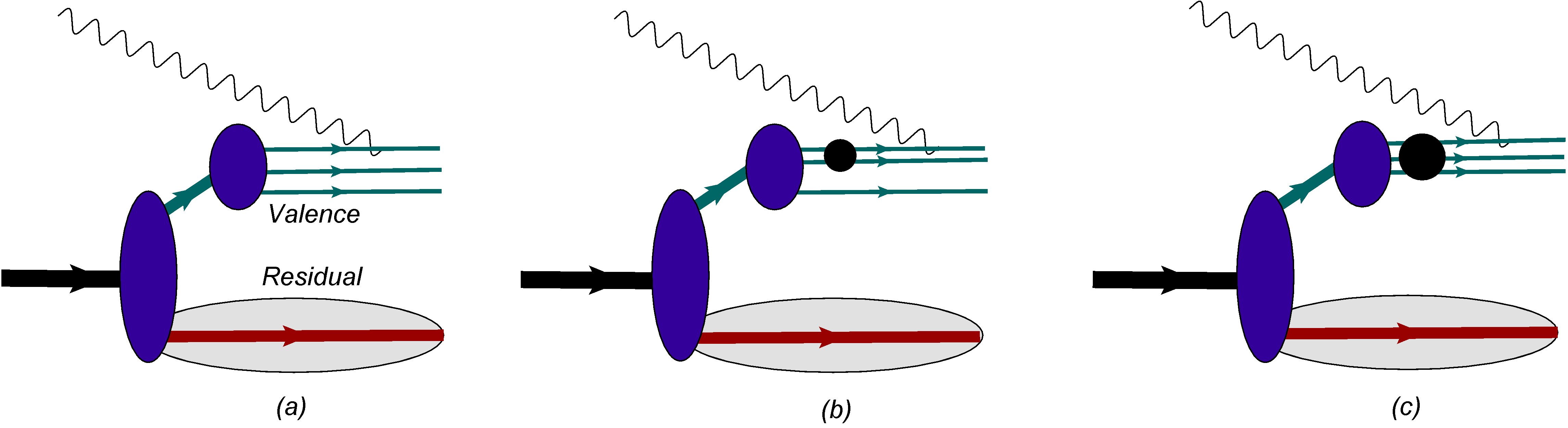}		
\centering
%\vspace{-0.5cm}
\caption{The mean field (a), two-valence (b)   and  three-valence (c) quark short-range correlation contributions to the deep-inelastic scattering 
off the nucleon in the partonic picture.}
\label{framework}
\end{figure*}

Such a framework in leading order, is represented by the three light-cone time-ordered  diagrams   in Fig.\ref{framework} and 
it is assumed that they should describe the major properties of  partonic  distribution at Bjorken $x>0.1$. For the smaller x the valence 
quark distributions are defined predominantly  by Regge dynamics dominated by $\alpha = {1\over 2}$ trajectory, which is beyond the scope of this paper 
(see e.g. Ref.\cite{Roberts:1990ww}).

Before proceeding with the calculations within this framework,  one can evaluate the  valence quark  distributions parametrically to validate the model at least on a 
qualitative level.  First,  the mean-field dynamics is largely responsible for the bulk properties of the bound three-quark system
such as its size and average momentum of its constituents. The functional form of the latter  can be described by an 
exponential function with the exponent being proportional to the spatial extension of the valence quarks.

\begin{figure*}
\includegraphics[width=1.0\textwidth]{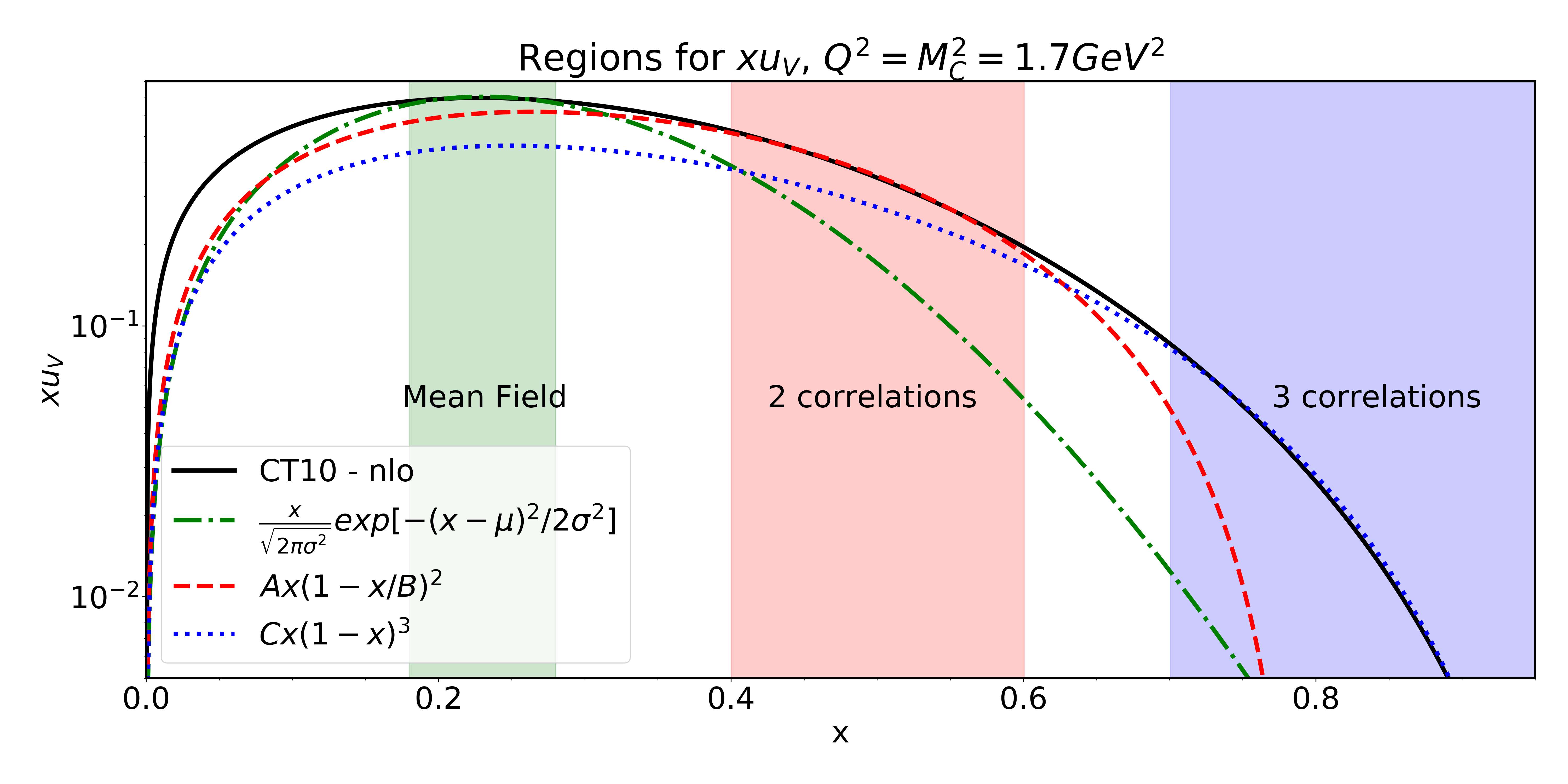}	
\centering
%\vspace{-0.5cm}
\caption{The graph shows the comparison of the x weighted valence u-quark distribution fitted by  three analytic functions corresponding to 
the mean field,  2q- and  3q- correlations.}
\label{xu_vs_framework}
\end{figure*}
   
To identify the range of Bjorken $x$ that one expects to be 
generated by mean field dynamics we note  that the phenomenology of PDFs is well consistent 
with  the average transverse momentum of  valence quarks  being about  $\langle k_t\rangle\sim 200$~MeV/c\cite{Feynman:1973xc}. 
Using the approximate spherical symmetry of  momentum distribution, we estimate that the Bjorken $x$ range relevant to the mean-field to be 
$x\lesssim  {2\langle k_t\rangle\over m_N} \sim 0.4$.  Thus for massless valence quarks one expects an exponential type distribution for the range of 
$0.1 \lesssim  x\lesssim  0.4$.

For the highest end of the Bjorken x range, the phenomenological observation is that for $x\ge 0.8$ the valence PDF's behave as $(1-x)^3$ which is 
consistent with the dynamics of two hard gluon exchanges between three valence quarks\cite{Brodsky:1974vy,Lepage:1980fj, Ball:2016spl} .  
Hereafter we refer them as three-quark (3q) correlations.
The short range nature of such a  correlation will require that all three quarks have momenta exceeding the above mentioned $\langle k_t\rangle$ and the interacting 
quark balances the other two spectator quarks. The latter results in $x \gtrsim {4\langle k_t\rangle\over m_n}\sim 0.8$.

Establishing the regions where one expects the dominance of mean-field and 3q-correlations our next conjecture is that the transition between these two 
regions happens through the two-quark short range interactions which we will refer to  as 2q-correlation.
%The above consideration allows  us to evaluate also  the region in $x$,  where one expects the dominance of two-fermion 
%correlations (if they exist)  as:  $x\sim {2\langle k_\perp\rangle\over m_N}\sim 0.4$ 
%(see e.g \cite{Frankfurt:1988nt}).  
If such two-quark correlations are due to short range  gluon  
interactions then they are dominated among two quarks with opposite helicites resulting in a functional form of the partonic distribution 
$(1-x/B)^2$, where $B$ is a parameter characterizing the total momentum fraction carried  by the center of mass of 
the 2q-correlated  pair.   

%With the same logic we estimate that three-quark, (3q)  correlations start to dominate when all three valence quarks have momenta exceeding average 
%transverse momenta, thus $x \gtrsim {3\langle k_\perp\rangle\over m_N}\sim 0.6$. If  3q correlations are generated by short range 
%vector exchanges the corresponding partonic distributions will have the well known analytic form of $(1-x)^3$.

In Fig.\ref{xu_vs_framework} we fitted 
 functional forms following from the  above discussed  scenarios of mean-field, 2q- and 3q- correlations to the  $x$ weighted
valence $u$-quark distribution. As the figure shows the exponential mean field as well as  $(1-{x\over B})^2$  and  $(1-x)^3$ forms for 
$2q$- and $3q$- correlations reproduce the shape of the valence quark distribution surprisingly well starting at $x\gtrsim 0.1$.
Note that in principle the same functions  should reproduce the $x<0.1$ part of the distributions, however, as it was mentioned earlier,
starting at $x<0.1$ valence quark  distributions are dominated by Regge behavior which is not included in the framework considered here.

\subsection{Phenomenological properties of valence quark  distributions}
\begin{figure*}[ht]
\centering
\includegraphics[width=1.0\textwidth]{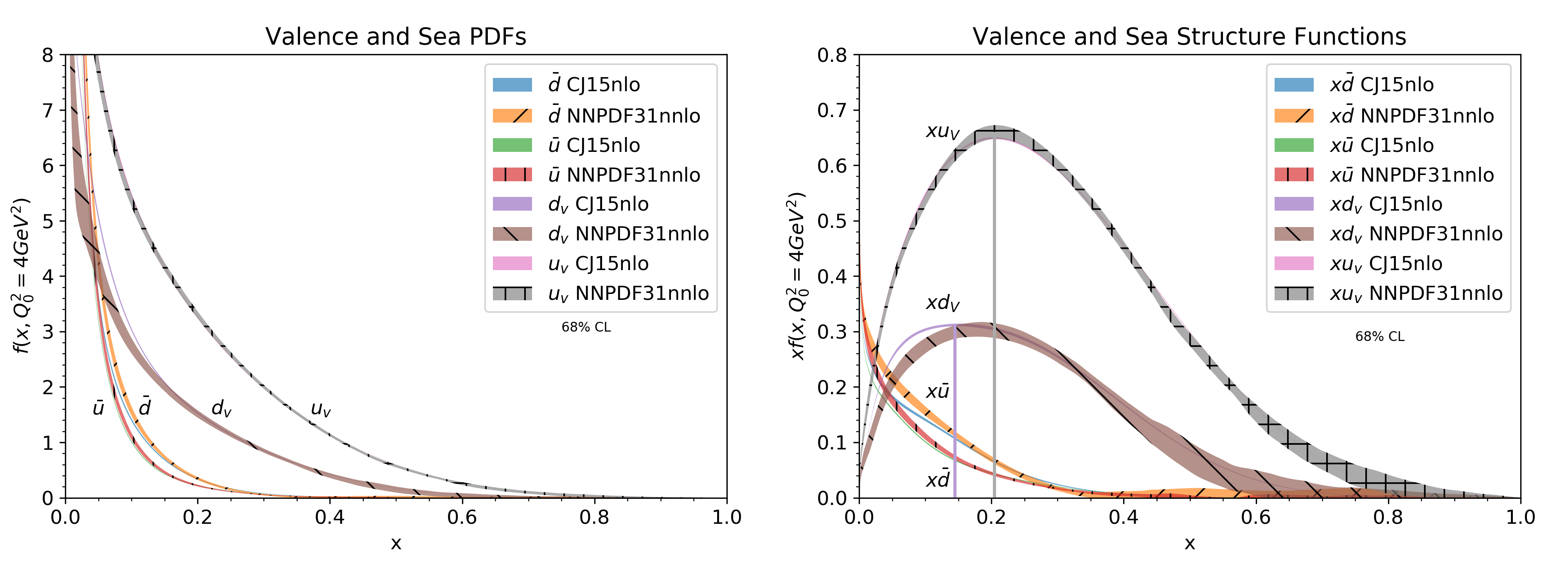}
\vspace{-0.5cm}
\caption{The up and down valence PDFs (left) and x weighted PDFs (right) at $Q^2 = 4 GeV^2$. 
The 68 \% confidence level error bands for PDF parametrization are shown.
The PDF sets CJ15nlo and NNPDF3.1-nnlo \cite{Accardi:2016qay,Ball:2017nwa} are used in the calculation. Similar features are also observed for other modern PDF parameterizations.}
\label{peak_and_height}
\end{figure*}

One of the most distinguished properties of valence quarks is that while their distributions do not exhibit any peaking structure,
their $x$  weighted distribution, $h(x,t)\equiv x q_V(x,t)$ has a 
clear peak  at around $x\sim 0.2$ which makes them 
qualitatively different from the sea quark distributions (Fig. \ref{peak_and_height}). 
Note that the peaking property  is universal for leading and higher order approximation despite the shape of PDFs not being a 
physical observable with certain degree of arbitrariness in their definition. Here we define $t = \log{Q^2}$. 

\begin{figure*}[ht]
\centering
\includegraphics[width=1.0\textwidth]{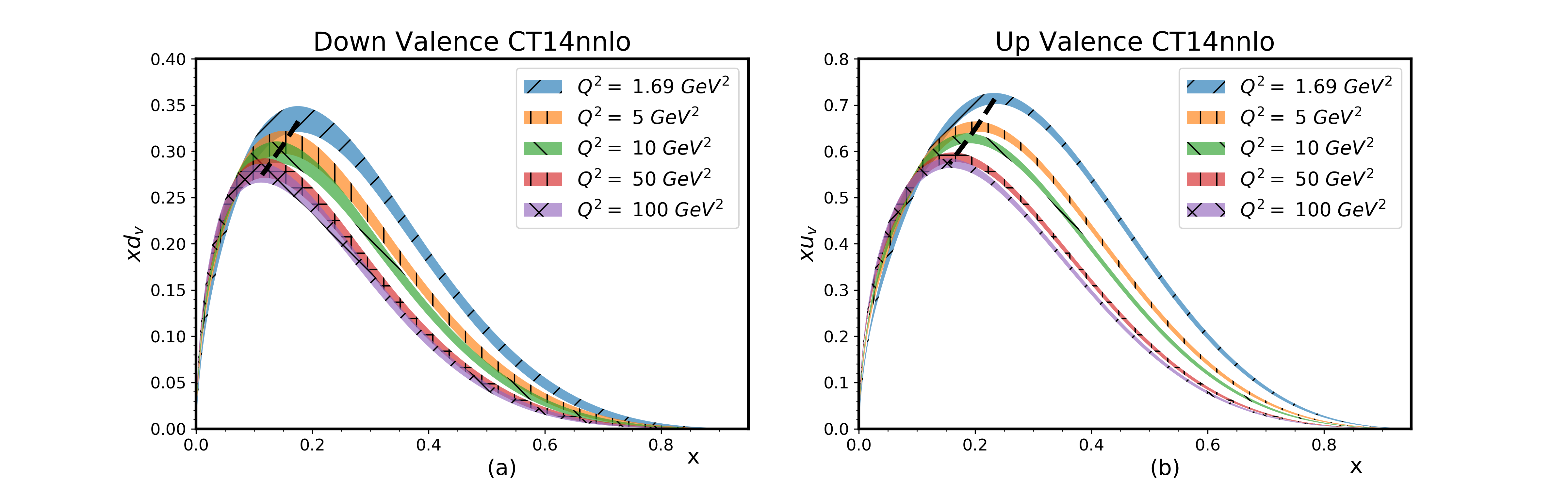}	
\vspace{-0.7cm}
\caption{The $x$ dependence of $x q_V(x,Q^2)$ function  for $d-$ and $u-$ valence quarks, calculated for  
CT14nnlo \cite{Dulat:2015mca} PDFs at various $Q^2$.  }
\label{peak_and_height_Q2dep}
\end{figure*}

One expects that the $Q^2$ dependence of the position and the height of the peak originate 
from QCD evolution, in which case with an increase of $Q^2$ the gluon radiations of valence quarks shift the peak towards 
smaller  $x$ and diminishes  the height  of the maximum. This can be seen in Fig.\ref{peak_and_height_Q2dep} where the valence 
$d$- and $u$- quarks structure functions are presented for $Q^2$ ranging from $5$ to $100$~GeV$^2$.

\begin{figure*}[thb]
\centering
\includegraphics[width=1.0\textwidth]{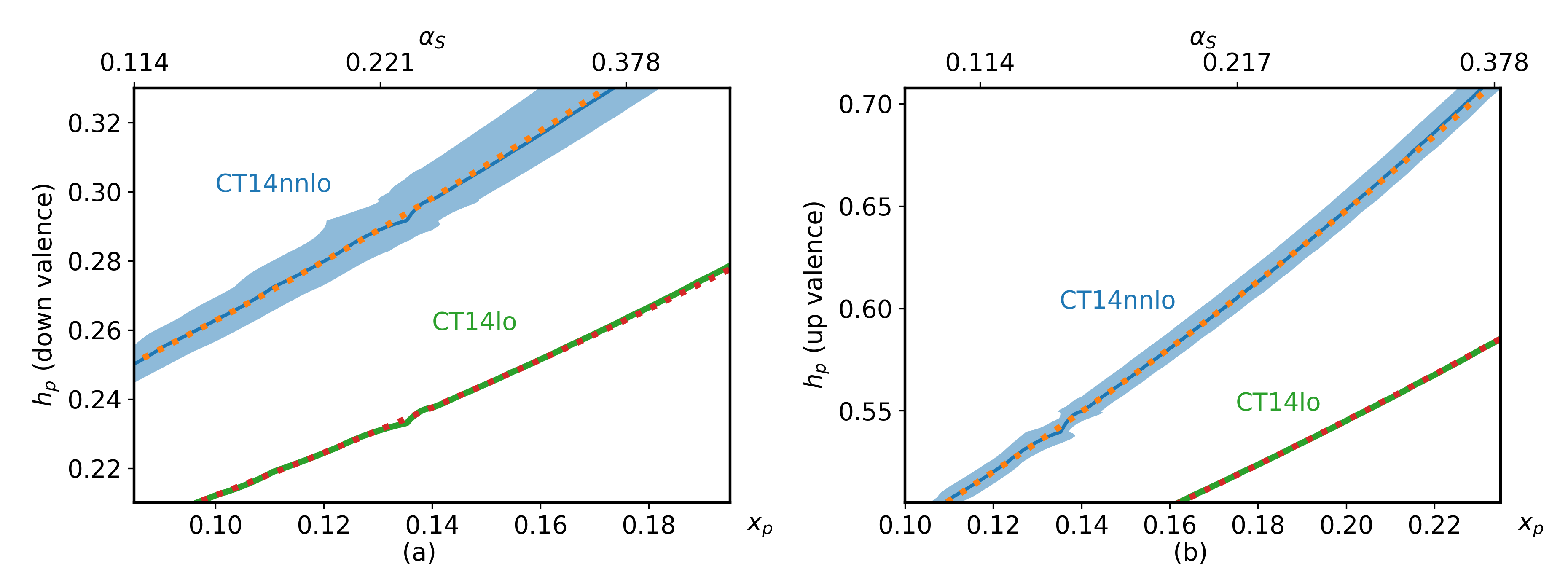}	
\vspace{-0.4cm}
\caption{The peak height and position correlation described by an exponential function: $h(x_p,t) =Ce^{Dx_p}$, 
from Ref.\cite{Leon:2020nfb}. CT14lo and CT14nnlo corresponds to the CT14 valence PDF parameterization
in leading and next to next to leading order approximation \cite{Dulat:2015mca}. 
Other PDF parameterizations also result in an exponential form of the peak height-position correlation. The CT14nnlo is the Hessian error at 68\% confidence level, while the CT14lo shows just the central value.}
\label{expcor}
\end{figure*}

In the recent work \cite{Leon:2020nfb} we found that the correlation between the height of 
the maximum of the $h(x_p,t)$ function  and its position, $x_p$  has a universal exponential form: $h(x_p,t) = C e^{Dx_p}$,
where constants, C and D saturate with the increase of the order of approximation in coupling constant.
As it can bee seen from  Fig.\ref{expcor} the exponential behavior extends to practically 
the entire range of the coupling constant $\alpha$ being covered in various deep-inelastic processes.

\begin{figure*}[ht]
\centering
\includegraphics[width=1.0\textwidth]{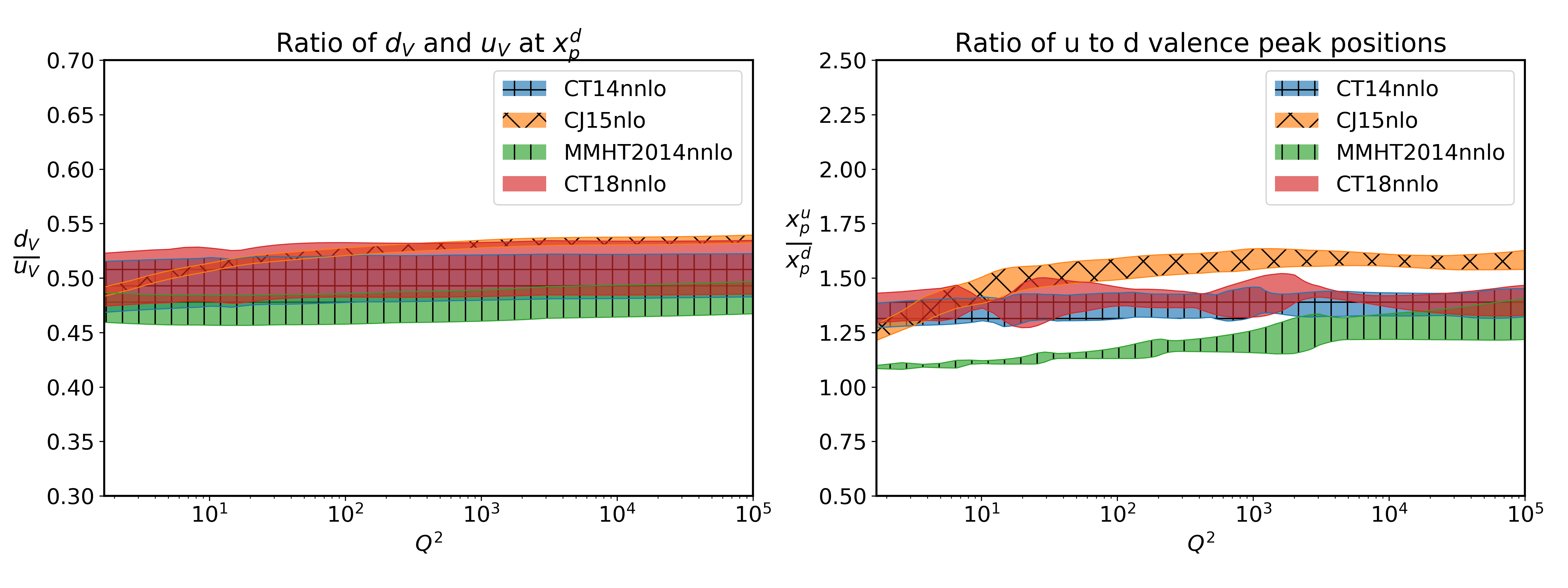}	
\vspace{-0.5cm}
\caption{(a) The $Q^2$ dependence of the ratio of d- to u- valence quark distributions at the peak position of $h(x,t)$ distribution for $d$ quark. 
(b) The $Q^2$ dependence of the ratio of the peak positions of the d- and u- valence quark $h(x,t)$ distributions.
Different PDF sets are used from Refs.\cite{Accardi:2016qay,Martin:2009iq,Dulat:2015mca,Owens:2012bv,Harland-Lang:2014zoa} .}
\label{R_Q2deps}
\end{figure*}

The observed ``height-position" correlation results in the following 
analytic form of the valence PDFs at the vicinity of $x_p$\cite{Leon:2020nfb} :
\begin{equation}
h(x,t) \equiv x q_V(x,t)  \approx  C + CDe x(1-x)^{1-x_p(t) \over x_p(t)}.
\label{meanfield}
\end{equation}
It can be checked that for  the above relation the $h(x,t)$ function  peaks at $x =x_p$. 
Eq.(\ref{meanfield}) indicates that the exponent of the $(1-x)$ term, ${1-x_p(t) \over x_p(t)}$, is defined by the position of the peak $x_p$, 
which changes continuously with $t$ due to QCD evolution. 
However as Fig.\ref{peak_and_height_Q2dep} (see also Fig.(\ref{R_Q2deps}(b) below) shows,  already at starting $Q_0^2=1.69$~GeV$^2$ 
(before the onset of QCD evolution) the peak-positions $x_p$ are different for valence d and u quarks, resulting in different $x$-dependencies of PDFs at the vicinity of  peaks. The observation that the exponent of  $(1-x)$ term is flavor-dependent indicates a more complex dynamics in the generation of  valence PDFs at $x\sim x_p$ than one expects, for example,  from the mechanism of a fixed number gluon exchanges that will 
result in a same exponent for valence u- and d- quark PDFs. A flavor-dependent  exponent can be  generated by considering an effective  
nonperturbative potential in Weinberg type equations for relativistic bound states \cite{Weinberg:1966jm}.

\begin{figure*}[ht]
\centering
\includegraphics[width = 0.8\textwidth]{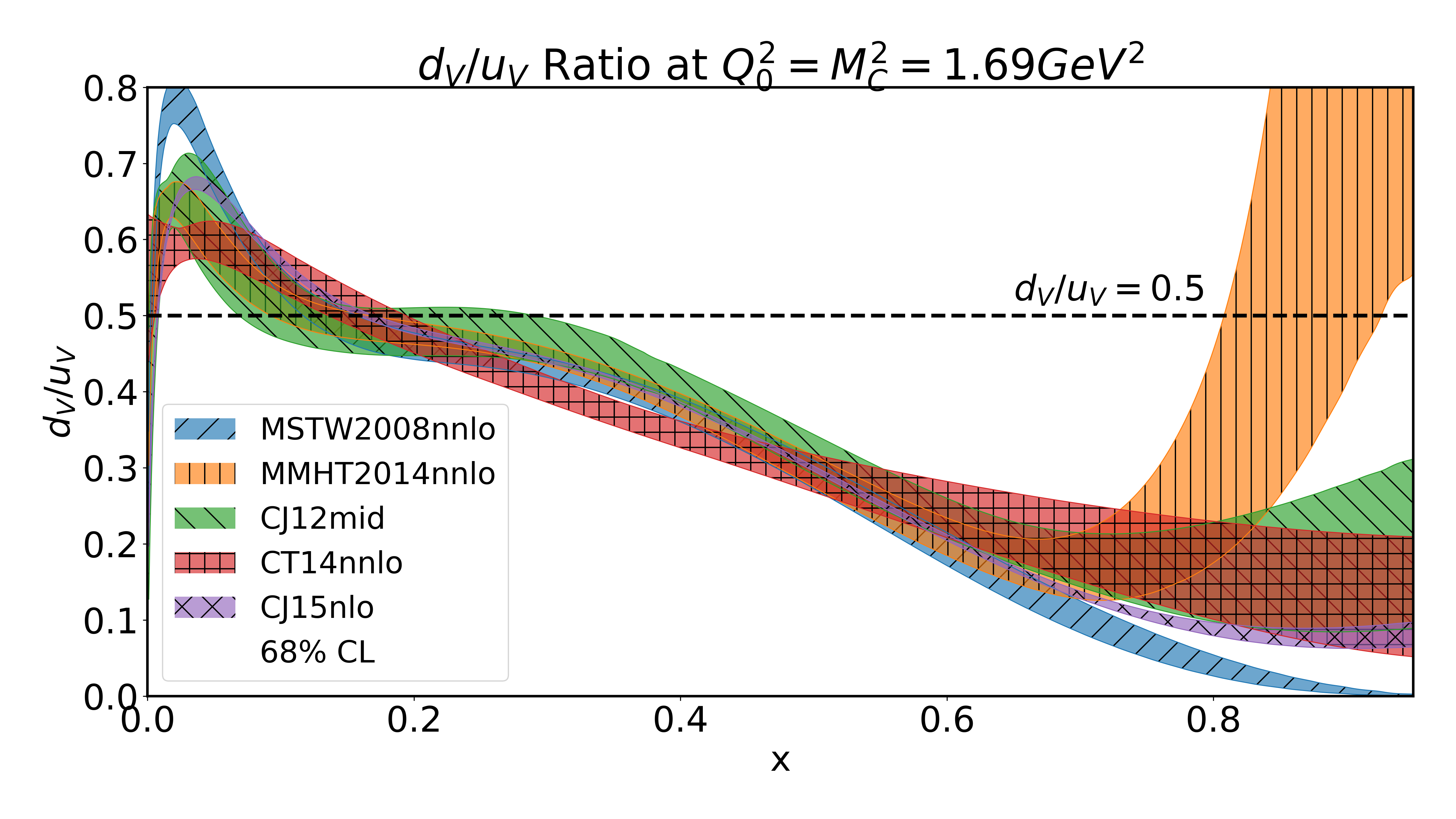}	
\vspace{-0.4cm}
\caption{The ratio of the $d-$ to $u-$ valence  quark distributions for various PDF sets \cite{Accardi:2016qay,Martin:2009iq,Dulat:2015mca,Owens:2012bv,Harland-Lang:2014zoa}
calculated at the lowest point of $Q_0^2 = M_c^2 = 1.69$~GeV$^2$ for which all the PDFs are defined.}
\label{dv_uv_ratio}
\end{figure*}

Since  the mean-field part  of the valence quark wave function will dominate in the calculations related to the bulk (integrated)  characteristics 
of baryons  (e.g.  mass spectrum \cite{Isgur:1979be}), then to explain the approximate SU(6) symmetry of baryonic spectroscopy one should 
expect the same symmetry to hold for the mean-field part of the valence PDFs. For SU(6) symmetry one expects 
that the ratio of valence  $d$- to $u$- quark distributions to scale as 0.5.  As Fig.\ref{dv_uv_ratio} shows all phenomenological valence quark distributions 
produce a $d_V/u_V$ ratio close to $0.5$ in the region, $x\sim x_p \sim 0.2$,  where one expects the dominance of mean-field dynamics. 
It is quite intriguing that some of these distributions such as CJ12\cite{Owens:2012bv} and 
MMHT2014\cite{Harland-Lang:2014zoa} produce an almost  constant scaling for $d_V/u_V$ ratios close to 0.5 at $x\approx x_p$
\footnote{It will be interesting if in future  analyses of the partonic distributions special attention will be given  to the unambiguous 
verification of the existence of  0.5 scaling  in $d_V/u_V$  ratios in $x\sim x_p \sim 0.2$  region.}.
The figure also shows the decrease of the $d_V/u_V$  ratio with an increase of x ($>$0.3), which is consistent with 
the  increasingly large  violation of SU(6) symmetry.  
One such mechanism that can result in such a violation is the onset of  quark-correlation dynamics in which  case one expects the violation of SU(6) symmetry due to the  helicity selection rule in quark-quark interactions through the hard vector~(gluon) exchanges (see e.g. Ref.\cite{Lepage:1980fj}).

In  Fig.\ref{R_Q2deps}(a) we show the  $Q^2$ dependence of the $d_V/u_V$ ratio estimated at the peak position of the 
$h(x,t) = x\cdot d_V(x,Q^2)$ distribution.  
As the figure shows the ratio $\approx 0.5$ is largely 
$Q^2$  independent, which indicates that such a ratio reflects an underlying property of the quark wave function of 
the nucleon  rather than being due to QCD evolution.  The  QCD evolution only  slightly modifies this ratio due to the migration of 
possible quark correlation effects from large to small  x region.

Finally, another phenomenological  feature is that for valence d- quarks the position of the peak 
of the $h(x,t)$ function is systematically lower than that of the valence u- quarks 
(see Fig. \ref{peak_and_height} (b)).  If such a difference in the peak positions is due to the dynamical property of 
valence d- and u- quarks in the nucleon, then one expects that it should be  insensitive to QCD evolution and be largely 
independent on $Q^2$.
Such an expectation is justified by the weak $Q^2$ dependence of the ratio of peak-positions of valence u- and 
d-quark distributions  for CT class of PDF parameterizations shown in Fig.\ref{R_Q2deps} (b).
The CJ15 parameterization shows $Q^2$ independence  starting at $Q^2\approx 10$~GeV$^2$ while
the MMHT14 PDFs reach $Q^2$ independence at $Q^2 > 10^3$~GeV$^2$.  
Even though the magnitude of the gap between peaks of $xd_V(x,t)$ and $xu_V(x,t)$  varies 
noticeably  between different groups of PDFs its $Q^2$ dependence is largely flat, not showing any systematic
$Q^2$ dependence.

 Summarizing this section we would like to emphasize the dedicated study of the above discussed characteristics of valence quark 
 PDFs, such as  exponential correlation of Fig.\ref{expcor}, the approximate $0.5$ scaling of d- to u- valence quark PDF ratios
 at $x\sim x_p\sim 0.2$ (Figs.\ref{dv_uv_ratio} and \ref{R_Q2deps}(a)) as well as the gap between d- and u- valence quark 
 peaking positions will  give new venues  in understanding the valence quark structure of the nucleon.
 Finally, even though in above discussion we considered PDF in higher order approximation (nlo and nnlo), 
 the observed PDF characteristics are largely the same also for leading order PDFs.

\subsection{Main assumptions of the model}

The above discussed phenomenology of valence PDFs  serves us as a ground for several assumptions for the model  outlined below:  

 \begin{itemize}
 \item {\bf Dynamics:} The main assumption of the model is that the  mean field, two- and three- quark short-range correlations define the dynamics of the valence quarks   in the range of  $0.1 \lesssim x\lesssim 1$. In the considered model, lepton scattering from the valence quark, 
 in leading order,  proceeds according to the  diagrams presented in Fig.\ref{framework}.
 
 \item {\bf Valence Quarks in the Nucleon:} The model assumes an existence of  almost massless three-quark   valence cluster, $V$, in the nucleon. 
 The cluster   is compact with the transverse separation between any pairs of quarks  $b_{qq} \lesssim 0.3$~Fm which indicates that 
 the 3q system will occupy a size of up to 0.6Fm.    Such a separation is 
 consistent with both phenomenology and theoretical evaluations made in all cases in which the 3q core of the nucleon 
 is considered (see e.g.\cite{Brodsky:1985qs,Cheedket:2002ik,Weiner:1985ih,Islam:2004ke}).
 In the model the 
 valence quark system defines the baryonic number, but not necessarily  the  total isospin  of the nucleon. It can have total isospin 
 $I_V = {1\over 2}$ or ${3\over 2}$,   each of them corresponding to  different excitations or masses of the residual nucleon system. 
 (For the lowest mass of the recoil system one expects  the 3q system to have the same isospin and its projection that 
  the considered  nucleon has).
  
 \item {\bf Residual Structure:} Introducing the residual structure of the nucleon with the spectrum of mass, $m_R$,  represents an 
 introduction of spectral function formalism in the description of the nucleon structure. 
 The model assumes a certain universality of the  residual structure, $R$,  entering in all three mechanisms of generation of 
 valence quark distribution according to Fig.\ref{framework}.  This universality is reflected in the fact that one can fix its main properties within one of the approaches (say within mean field model) and apply  it in the calculation of $2q$- and $3q$- short range correlation contributions.  
The mass spectrum of the residual system is continuous  and effectively depends on whether u- or d- valence quarks are probed.
For example, a  given valence quark  in the proton, can originate from the 3-valence quark cluster having the total isospin and its projection the same as 
the proton but it can also originate from the 3-quark cluster with isospin, $I_V={1\over 2}$ and projection $I_V^3=-{1\over 2}$
or isospin,  $I_V={3\over 2}$, corresponding to the higher mass 
of the residual system. Thus 
in the case of the proton one can expand the residual nucleon mass in the form:
\begin{eqnarray}
m_R(u/d) = \alpha_{u/d}\cdot  m_R(I_V={1\over 2},I^3_V= {1\over 2} ) \\+  \beta_{u/d} \cdot m_R(I_V={1\over 2},I^3_V= -{1\over 2} ) \nonumber \\
+ \gamma_{(u/d)} \sum\limits_{I^3_V=-{3\over 2}}^{3\over 2} m_R(I_V={3\over 2},I^3_V ) + \cdots \nonumber
\label{massspectrum}
\end{eqnarray}
where $I^3_V$ is the isospin projection of the valence quark cluster and ``$\cdots$" accounts for the possible 
higher mass spectrum of the residual system. 
Here the $\alpha$, $\beta$ and $\gamma$  factors are estimating the probabilities that 
u- or the d- valence quarks emerge from different isospin configurations.

From the above representations 
one expects that the minimal mass in the residual system comes from the contribution 
of $I_V={1\over 2},I^3_V= {1\over 2}$ term that has the same isospin  numbers that the proton has.
As it was mentioned above, for the cases of total isospin ($I_V={1\over 2}$, $I_V^3=-{1\over 2}$) or   $I_V={3\over 2}$ the residual system will correspond to higher mass excitations.
With this assumption, we observe  that, for the proton having $uud$ configuration, (per one valence quark)  the  u- valence quark  will  be accompanied with lesser residual mass than that of the  d-quark.   Thus one expects 
that for proton in general:
\begin{equation}
m_R(u) < m_R(d),
\label{udmass}
\end{equation}
since for the d- valence quarks one expects relatively larger contribution from higher residual masses.
For example, the second $\beta$ term contributes more in the d- valence quark case since it has $udd$ configurations.
It is interesting that such a scenario is in qualitative agreement  with violation of Gottfried sum rule\cite{Arneodo:1994sh} or the 
"SeaQuest"  result\cite{Dove:2021ejl}, since large $\beta$ factor for d- valence quark will generate an excess of $\bar d$ quarks  
due to $(ddu)(\bar du)$ configuration in the proton.

In the current paper we will estimate the magnitude  of $m_R$ for u- and d- quarks by fitting 
the calculated valence quark distributions to the empirical PDFs.
In principle it could be  possible to extract $m_R$  experimentally, with dedicated measurements of   
tagged structure functions   in semi-inclusive DIS processes. 

\item {\bf Mean Field Model:} In the considered model we assume that the mean-field dynamics (Fig.\ref{framework}(a)) 
is largely responsible for the position and the peak of the valence quark structure  function, $h(x,t)$.
We assume that mean-field dynamics preserves the SU(6) flavor-spin symmetry  of valence quark-cluster for both the $I_V = {1\over 2}$  and ${3\over 2}$ cases. Expecting that mean the field 
contribution will be relevant for the region of  x up to as much as 0.55-0.65  and requiring 
the produced final mass in DIS is $\ge 2$~GeV, we  set the starting  $Q^2 = 4$~GeV$^2$.

\item{\bf $2q$- and $3q$- Correlations:} Starting at $x\gtrsim 0.4$ the model assumes the onset  of $2q$- short range correlations transitioning to $3q$ correlations 
at $x\gtrsim 0.7$.  All such correlations are taken place within the valence quark cluster.
We assume that such short range correlations are generated by vector exchanges, which in the pQCD limit will correspond to the hard gluon exchanges.
In the calculation of such interactions the finite range of correlations will be introduced by the momentum transfer cut-off in the gluon exchanged diagrams.
Because of the selection rule, in which  gluon exchanges are dominated for quarks with opposite helicities in the massless limit, we expect that with the onset of 
quark correlations the SU(6) symmetry of the 3q system will break down. 
 \end{itemize}

Using the above assumptions in developing the quark correlation model of valence quark distributions we  also 
bring forward several experimentally verifiable predictions. These include the prediction for universality of the recoil system, $R$, for 
almost entire range of $x>0.1$ as well as  the prediction for the onset of $2q$- and $3q$- short range quark correlations which 
will be distinguished by specific correlation properties  between struck and recoil valence quarks. 
These predictions can be quantified once the model parameters are 
fixed by comparing the calculated valence quark distributions with PDFs extracted from DIS measurements.

It is worth mentioning that the presented approach can be applied also for the studies of QCD processes in the 
nuclear medium\cite{Sargsian:2002wc,Cosyn:2017ekf,Cosyn:2010ux},
in which  case an additional vertex will be added to the diagrams of Fig.\ref{framework} describing the transition of nuclear target 
to the ``nucleon- residual nucleus"  intermediate state.

\section{Mean-Field Model of Valence Quark Distributions}
\label{model}

Our focus now is on modeling the mean-field dynamics of valence quark interaction and 
calculating the valence quark distribution functions that can be compared with empirical
PDFs.   The mean field picture that we adopt assumes that the nucleon core 
consists of interacting thee-valence quark cluster embedded in the residual nucleon system 
consisting of sea quarks, gluons, pions and other possible hadrons.  
The main assumptions of the mean field model  are as follows:
\begin{itemize}
\item The valence 3q system occupies a space in which the separation of 
any given quark pair in the impact  parameter space is about $0.3$~Fm.
The interactions among these quarks is described by  coupled relativistic
three-dimensional harmonic oscillators, thus satisfying confinement condition. 
\item Valence quarks are almost massless  with the invariant energy of the 3q system contributing 
to the  nucleon mass.
\item The residual  system generates an external field for 3q valence system and occupies  a volume  less or equal to  the nucleon volume.
\end{itemize}

\begin{figure}[ht]
\includegraphics[width=\columnwidth]{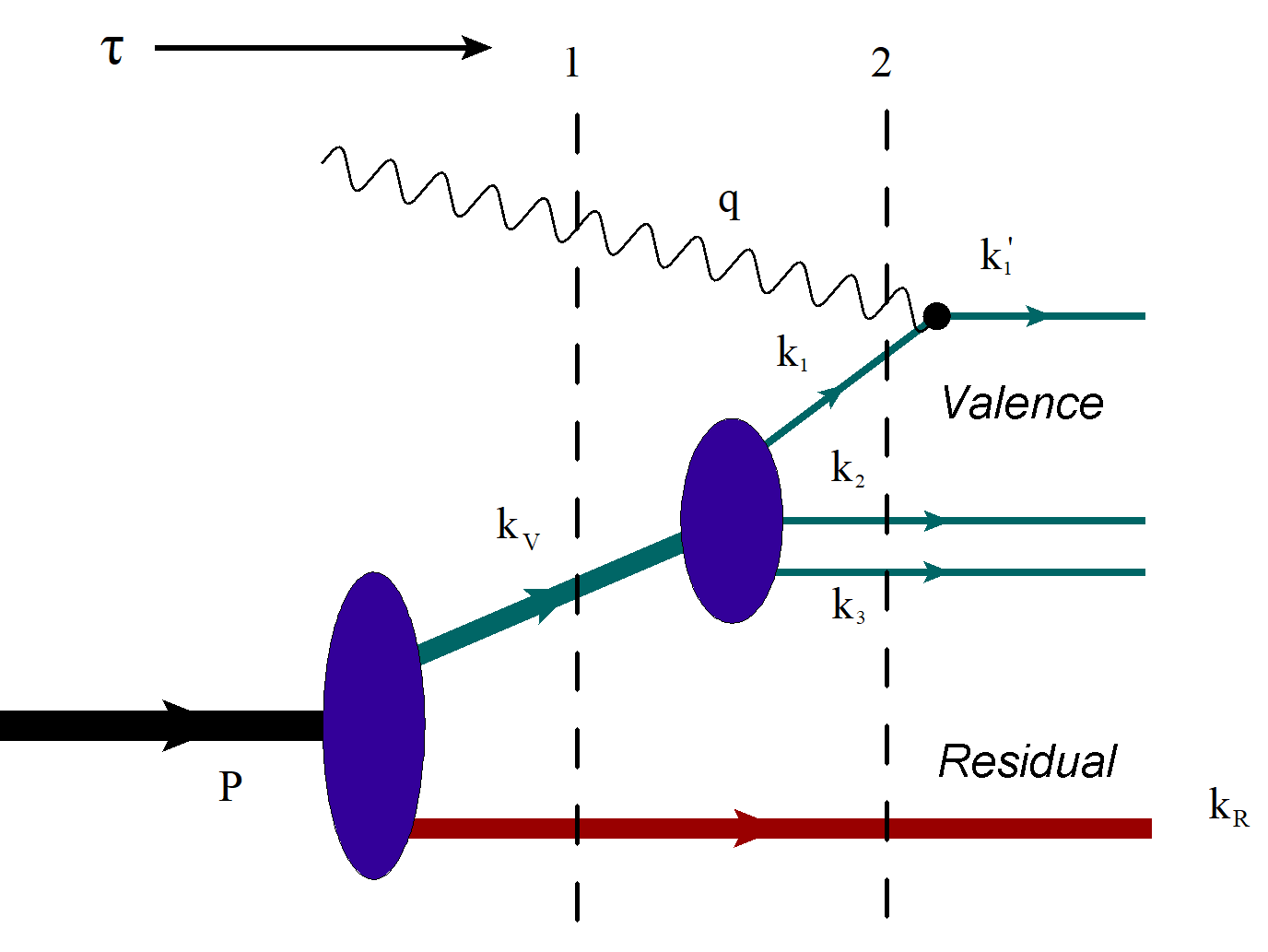} 
\centering
\vspace{-0.4cm}
\caption{The nucleon transition into a valence and residual systems. 
The diagram is arranged by light-front
$\tau = x^+ = z + t$ ordering.  The two dashed vertical lines represent the two intermediate states 
where the light-front wave functions of residual and 3q systems are defined.}
\label{mf_diagram_lc}
\end{figure}

To calculate the amplitude of virtual photon interaction with the valence quark in the above described system we  
notice that,  in the leading order, the  light-front time, $\tau$ evolution of scattering process proceeds according to 
the diagram in Fig.\ref{mf_diagram_lc}. Here the first vertex (from the left) describes the interaction of 
the  3q valence quark-cluster  in the field of the residual system and the second vertex, 
the interaction among three valence quarks.

The scattering amplitude corresponding to the diagram of Fig.\ref{mf_diagram_lc} can 
be calculated by applying effective light-front perturbation theory in which one introduces 
effective vertices that can be absorbed into the definition of light-cone wave 
functions. The present approach represents the light-front projection of the covariant amplitude 
which  formally can be calculated within the effective Feynman diagrammatic  
approach\cite{Sargsian:2001ax}.

\subsection{General Considerations: Reference Frame, Kinematics and the Cross Section of the Reaction.}

The calculations here will be done using light-cone~(LC) coordinates with the 4-momenta and 
four-products defined  as follows:
\begin{eqnarray}
& k^\mu = (k^+, k^-, \mathbf{k}_\perp),  \ \   k^\pm = E \pm k_z, \ \ \mathbf{k}_\perp = (k_x, k_y), \ \ \nonumber \\ 
&k_1 \cdot k_2 = {k_{1}^-k_{2}^+\over 2} + {k_{1}^+k_{2}^-\over 2} - {\bf k_{1, \perp}\cdot k_{2, \perp}}.
\end{eqnarray} 
We consider a reference frame where the four-momenta of the nucleon, $p^\mu_N$ and the virtual photon, $q^\mu$  are:
\begin{eqnarray}
& p_N^\mu =  (p^+_N, \frac{m_N^2}{p^+_N}, \mathbf{0}_\perp), \ \  
q^\mu = (0, \frac{2p \cdot q}{p^+_N}, \mathbf{q}_\perp), \ \  \nonumber  \\
& Q^2 = - q^2 =  |\mathbf{q}_\perp|^2,
\label{reframe}
\end{eqnarray}
where $m_N$ is the mass of the nucleon and we use the standard definition of Bjorken variable: $x_B = \frac{Q^2}{2 p \cdot q}$. 

The important kinematical condition of the chosen reference frame is that  we assume: 
\begin{equation}
p_{N}^+ \gg m_N, k_{i}^-,k_{i,\perp},
\label{highkin}
\end{equation}
where $k_i^-$  and $k_{i,\perp}$ are the ``-" and transverse  components of the momenta of constituents in the nucleon.

In calculating the cross section of deep inelastic inclusive scattering from the nucleon 
it is convenient to introduce the nucleonic tensor, $W^{\mu\nu}$, through which in 
the one-photon exchange approximation  the 
differential cross section can be presented as follows:
\begin{equation}
{d\sigma\over dQ^2 dx} = 
{2\pi \alpha^2\over Q^4 }{y^2\over Q^2}m_NL_{\mu\nu}W_N^{\mu\nu},
\label{dsigma0}
\end{equation}
where  $y = {p_N \cdot q\over p_N \cdot k}$ and the spin averaged leptonic tensor is defined as:
\begin{equation}
L_{\mu\nu}  = 
4[(k_\mu - {q_\mu\over 2})(k_\nu - {q_\nu\over 2})] + Q^2[-g_{\mu\nu}- {q_\mu q_\nu\over Q^2}], 
\end{equation}
where $k_\mu$ is the four momentum of initial lepton.

If one introduces  the deep-inelastic nuclear transition current as $J_N^{\mu}(p_X,s_X,p_N,s_N)$ between the initial  nucleon and the  final deep inelastic  state, X,   then the 
nucleonic tensor can be presented as
\begin{align}
W_N^{\mu\nu} &=  {1\over 4\pi m_N }\int \sum\limits_X\sum\limits_{s_X}J^{\mu,\dagger}(p_X,s_X,p_N,s_N) \nonumber \\
&\times J^{\nu}(p_X,s_X,p_N,s_N)  
 (2\pi)^4\delta^4(q+p_N - p_X)  \nonumber \\ 
&\times \delta(p_X^2-M_X^2) {d^4p_X\over (2\pi)^3},
\label{Wmunu}
\end{align}
where one  sums  over the  all possible finite DIS states, $X$.  For the most general case the 
nucleonic tensor for unpolarized electron scattering from an unpolarized target can be expressed 
through two invariant structure functions 
\begin{align}
W_N^{\mu \nu} &= - {F_1 (x_B, Q^2)\over m_N} ( g^{\mu \nu} - 
\frac{q^\mu q^\nu}{q^2})  \nonumber \\
&+ 
\frac{F_2(x_B, Q^2)}{m_N  (p_N\cdot q)}(p_N^\mu - \ \frac{p_N \cdot q}{q^2} q^\mu )(p_N^\nu -\frac{p_N \cdot q}{q^2} q^\nu ),
\end{align}
where in the reference frame of Eq.(\ref{reframe}):
\begin{align}
F_{2}(x,Q^2)  & =  {m_N (p_N\cdot q)\over (p_N^+)^2} W_N^{++} = {m_N Q^2\over 2x (p_N^+)^2)}W_N^{++}
 \nonumber \\
F_{1}(x,Q^2) & =  {F_{2}(x,Q^2)\over 2x}\left(1 + {2m_N^2x^2\over Q^2}\right) \nonumber \\
&- {m_N\over 2} W_N^{+-},
\label{F2F1}
\end{align}
and because of the gauge invariance,
\begin{equation}
W_N^{+-} = {2Q\over q^-}W_N^{\perp-}.
\label{gaugeinv}
\end{equation}
With these structure functions the invariant cross section of the scattering (\ref{dsigma0}) can be presented 
as follows:
\begin{align}
 &{d\sigma \over dQ^2 dx}  =    
  {2\pi \alpha^2\over Q^4}y^2\bigg( 2F_{1}(x,Q^2) \nonumber \\
  &+ {1\over 2 x y^2}
\bigg[(2-y)^2 - y^2\bigg(1 + {4m_N^2 x^2\over Q^2}\bigg)\bigg]F_2(x,Q^2)\bigg). 
\label{dsigma}
\end{align}
With respect to Eq.(\ref{F2F1}) it is worth noting that in the limit of ${2m_N^2x^2\over Q^2}\ll 1$, if 
$W^{+-} = 0$, one arrives at Callan-Gross relation:
\begin{equation}
F_{2}(x,Q^2)  = 2 x F_{1}(x,Q^2).
\label{CGrel}
\end{equation}
This,  together with Eq.(\ref{gaugeinv}) indicates that the Callan-Gross relation is fulfilled when 
$W^{\perp-} = 0$, which according to Eq.(\ref{Wmunu}) corresponds to the absence of the transverse 
currents $J^\perp=0$ in the deep-inelastic scattering process in the chosen reference frame (\ref{reframe}).
 
In the partonic model  the structure function $F_2(x,Q^2)$ is directly related to the partonic distribution functions \cite{Feynman:1973xc}:  
\begin{equation}
F_2(x,Q^2) = \sum\limits_i e_i^2 x f_i(x,Q^2),
\label{partmodel}
\end{equation}
where the $e_i$'s are the charges of interacting quarks relative to the magnitude of electron charge and 
$f_i(x,Q^2)$ is the PDF of the quark of type $i$. 

From  Eqs.(\ref{F2F1}) and (\ref{partmodel}) it follows that to calculate the PDFs one needs to calculate the
$W^{++}$  component of the  nucleonic tensor, $W^{\mu\nu}$.

\subsection{Calculation of the Scattering Amplitude}
The transition current, $J^\mu$ in Eq.(\ref{Wmunu}) in our approach is identified 
with the scattering amplitude $A^\mu$, which corresponds to the  process described by 
the diagram  of Fig.\ref{mf_diagram_lc}.  To calculate this amplitude we apply the effective 
light-front diagrammatic rules (summarized in Appendix A) which results in:
\begin{align}
&A^\mu =  \sum_{h_V, h_1} \frac{1}{k_V^+} \frac{1}{k_1^+}
 \bar{u}(k_1',h_1')   (ie_1 \gamma^\mu)   u(k_1,h_1) \nonumber\\ 
& \times {\prod_{i=1}^{3} \bar{u}(k_i, h_i)\Gamma^{V \rightarrow 3q} \chi_V \over 
\mathcal{D}_2}
 {\chi^\dagger_V  \chi^\dagger_R   \Gamma^{B \rightarrow VR} u(p_N,h_N)\over \mathcal{D}_1}
 \label{Amu1}
\end{align}   
where  $\Gamma^{N \rightarrow VR}$ is the effective vertex for the 
nucleon transition to the valence quark cluster $V$ and residual $R$  systems, 
the $\Gamma^{V \rightarrow 3q}$ vertex describes the transition of 
the state V  into three individual valence quarks.
(These vertices are non-perturbative objects and  can not be calculated 
with  any finite order diagrammatic approach. In our framework they  will be associated with 
respective wave functions which will be modeled and evaluated by comparing with  experiment.)
All the $h$'s denote helicities, the  $\chi_V$  and $\chi_R$  denote the 
respective  spin functions for the valence  and residual systems
\footnote{For example, a spinor for a spin 1/2 configuration, a polarization vector for spin 1, etc.}. 
The light front energy denominators are $\mathcal{D}_1$ and $\mathcal{D}_2$,
which characterize two intermediate states marked by dashed vertical lines in Fig.\ref{mf_diagram_lc}.
Here we label the struck quark with $i=1$. The  color indices are omitted since they are not 
involved in electromagnetic interaction. 
The  sequence of $N\rightarrow VR$ followed by $V\rightarrow 3q$  transitions as presented in 
Fig.\ref{mf_diagram_lc} is justified based on the assumption that the $VR$ system is characterized by smaller internal momenta than the  $3q$ system.
% are characterized 
%by different internal momentum scales.
% $\sim {1\over r_{N}}$ and ${1\over 0.3fm}$ respectively.
The assumption of small relative momentum in the  $VR$ system allows us to approximate the first intermediate 
state as on-shell.  For the 3q system such an approximation is not valid and in principle for 
the valence quark propagator one will have an on-shell and instantaneous terms due to relation:
$\slashed{k} + m =  \sum_h u(k,h) \bar{u}(k,h) + \frac{\gamma^+(k^2-m^2)}{k^+}$.  However, 
since we are interested in the calculation of the $A^+$ component of the scattering amplitude 
which enters in $W^{++}$, the instantaneous 
term drops out since it couples with the extra $\gamma^+$ factor 
in the $\gamma q q$ vertex and $\gamma^+\gamma^+ = 0$. 
Thus, this allows us to represent the second intermediate state by on-shell quark spinors only.

Applying the LF diagrammatic rules (Appendix A) for $\mathcal{D}_1$ and $\mathcal{D}_2$ denominators one obtains:
\begin{align}
\mathcal{D}_1 &=   P^- - k_R^- - k_V^-\nonumber\\ 
&=  \frac{m_N^2}{p_N^+}  - \frac{k_{R, \perp}^2 +  m_R^2}{k_R^+}  - \frac{k_{V, \perp}^2 +  m_V^2}{k_V^+}) 
\nonumber \\
&=  \frac{1}{p_N^+} \left(m_N^2 - \frac{k_{R, \perp}^2 +  m_R^2}{x_R} - \frac{k_{V, \perp}^2 +  m_V^2}{x_V} \right) \nonumber \\
\mathcal{D}_2 & =  \frac{1}{k_V^+} \big(m_V^2 - \sum_{i=1}^{3} \frac{ k_{i, \perp}^2 +m_i^2 }{\beta_i} ), \label{D1D2}
\end{align}
where $k_R$ and $k_V$ are the four momenta of the residual and valence systems respectively, 
$m_R$ and $m_V$ are their masses and, the summation is  over the valence quarks, $i$,  
with $\beta_i = \frac{x_i}{x_V}$.

Substituting Eq.(\ref{D1D2}) into Eq.(\ref{Amu1}) results in:
%\begin{widetext}
\begin{align}
A^\mu &= \sum_{h_1, h_V} \frac{1}{x_V} \frac{1}{\beta_1}
   \bar{u}(k_1',h_1')   (ie_1 \gamma^\mu)   u(k_1,h_1) \nonumber\\
&\times \frac{  \prod_{i=1}^{3} \bar{u}(k_i, h_i)\Gamma^{V \rightarrow 3q} \chi_V}   
{m_V^2 - \sum_{i=1}^{3} \frac{ k_{ i, \perp}^2 +m_i^2 }{\beta_i} } \nonumber\\
&\times \frac{\bar{\chi_V}  \bar{\chi}_R   \Gamma^{B \rightarrow VR} u(p_N,h_N)}
{m_N^2 - \frac{k_{V, \perp}^2 + m_V^2}{x_V} - \frac{k_{R,\perp}^2 + m_R^2}{x_R}}.
\label{Amu2}
\end{align}
%\end{widetext}
To proceed,  we introduce   light-front wave  functions for  VR and 3q systems  as follows:
\begin{align}
& \psi_{VR} (x_V, \mathbf{k}_{R, \perp}, x_R, \mathbf{k}_{V, \perp}) \nonumber\\
&=   \frac{\bar{\chi_V  }\bar{\chi_R} \Gamma^{B \rightarrow VR} u(p_N,h_N)}{m_N^2 - \frac{k_{V, \perp}^2 + m_V^2}{x_V} - \frac{k_{R,\perp}^2 + m_R^2}{x_R}}  \nonumber \\ 
& \psi_{3q} (\{\beta_i, \mathbf{k}_{i, \perp}, h_i \}_{i=1}^3)  =     \frac{\prod_{i=1}^3 \bar{u}(k_i,h_i)  \Gamma^{V \rightarrow 3q} \chi_V  }
{m_V^2 - \sum_{i=1}^{3} \frac{ k_{ i, \perp}^2 +m_i^2 }{\beta_i} },
\label{LFwfs}
\end{align}
where $\{\beta_i, \mathbf{k}_{i, \perp}, h_i \}_{i=1}^3$  denotes the LC momenta and helicities of the three valence quarks in the wave function.

Using the above definitions of LF wave functions the scattering amplitude can be presented in the following form:
\begin{align}
A^\mu &= \sum_{h_1, h_V}\bar{u}(k_1,h_1)   (ie_1 \gamma^\mu)   u(k_1,h_1)  \nonumber  \\ & \times \frac{\psi_{VR}(x_V, \mathbf{k}_{R, \perp}, x_R, \mathbf{k}_{V, \perp})}{x_V} \frac{\psi_{3q}(\{\beta_i, \mathbf{k}_{i, \perp}, h_i \}_{i=1}^3) }{\beta_1}.
\label{Amu3}
\end{align}

\subsection{Calculation of  the Valence PDF}

\begin{figure}[ht]
\includegraphics[width= \columnwidth]{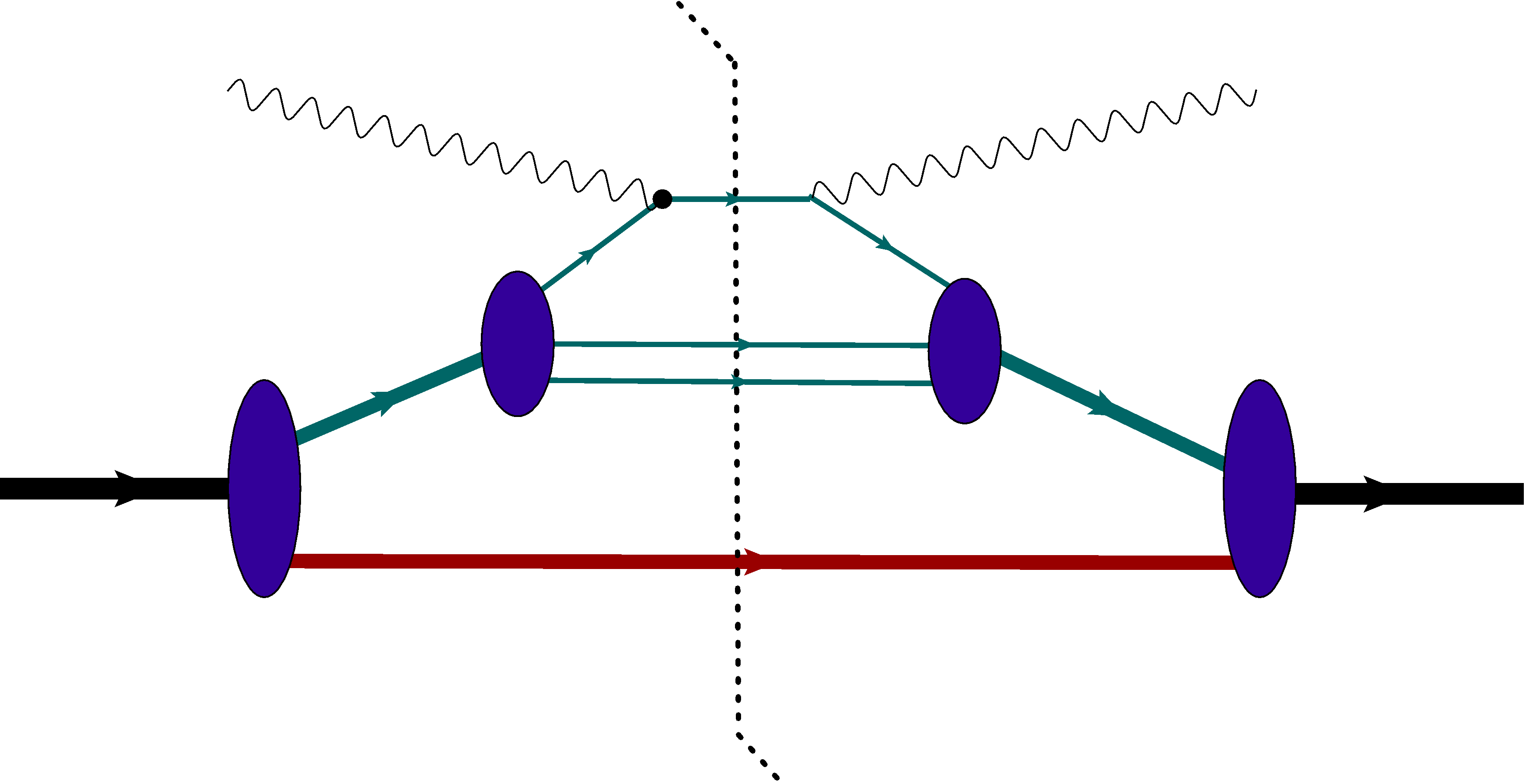}	
\centering
\vspace{-0.2cm}
\caption{Hadronic tensor expressed as a cut diagram.}
\label{cutdiagram}
\end{figure}

The hadronic tensor, $W_N^{\mu\nu}$ of  Eq.(\ref{Wmunu})  for the considered scenario of scattering  in which the final state 
consists of three outgoing valence quarks and a residual nucleon system, in the leading order,  corresponds to the 
cut diagram of Fig.\ref{cutdiagram} for which one obtains:
\begin{align}
&W_N^{\mu\nu} (x, Q^2)   =  \frac{1}{4\pi m_N} \sum_{ q, h_i} 
\int  \delta(k_R^2- m_R^2)   \frac{d^4 k_R}{(2\pi)^4} \nonumber \\
& \times \delta(k^{\prime,2}_1- m_1^2)   \frac{d^4 k^\prime_1}{(2\pi)^4} \prod_{i=2}^3 \delta(k_i^2- m_i^2)   \frac{d^4 k_i}{(2\pi)^4} \nonumber \\
& \times \delta^{(4)}(P+q - k^\prime_1+ \sum_{i=2}^3 k_i -k_R)  A^{\mu \dagger} A^\nu,
\label{Wmunumodel}
 \end{align}
where the sum $ \sum\limits_{ q, h_i} $ is over all the spins/flavors of the  outgoing valence quarks and the residual system.  The Lorentz invariant phase space of all final particles can be presented as follows: 
\begin{align}
\delta(k_j^2-m_j^2) d^4k &= \frac{1}{2} \delta(k_j^+k_j^- - k_{j,\perp}^2 -m_j^2) dk_j^- dk_j^+ d^2\mathbf{k}_{j, \perp} \nonumber \\
&=  \frac{ dx_j d^2\mathbf{k}_{j, \perp}}{2x_j} |_{k_j^- = \frac{k_{j,\perp}^2 +m_j^2}{x_jp_N^+}},
\label{onmass}
\end{align}
where $x_j =\frac{k_j^+}{p_N^+}$ is the light-cone momentum fraction of outgoing particles, $j=1^\prime,2,3$ and $R$.
To evaluate the energy-momentum conservation condition, first we express the $\delta^4()$ function through the LC components:
\begin{align}
& \delta^{(4)}(p_N+q -  k^\prime_1 - \sum_{i=2}^3 k_i -k_R) \nonumber \\
&=  2 \delta (p_N^+ + q^+ - k^{\prime,+}_1 - \sum_{i=2}^3 k_i^+ - k_R^+)  \nonumber \\
& \times \delta (p_N^- + q^- -  k^{\prime,-}_1 -  \sum_{i=2}^3 k_i^- - k_R^-) \nonumber \\
& \times \delta^{(2)}(\mathbf{p}_{N,\perp} + \mathbf{q}_\perp -  k^{\prime}_{1,\perp} - \sum_{i=2}^3 \mathbf{k}_{i, \perp} - \mathbf{k}_{R, \perp}).
\label{delta4}
\end{align}
Using the reference frame of Eq.(\ref{reframe}) in which $q^+=0$,  one observes that $k^+_1 = k^{\prime,+}_1$ and 
using the definition of  LC momentum fractions $x_j =\frac{k_j^+}{p_N^+}$  one obtains:
\begin{equation}
 \delta (p_N^+ + q^+ - k^{\prime +} _1- \sum_{i=2}^3 k_i^+ - k_R^+) = \frac{1}{p_N^+} \delta (1 - \sum\limits_{i=1}^3x_i- x_R).
\label{momplus}
 \end{equation}

For the ``-" component we use $q^- = {2p_N \cdot q\over p_N^+}$ (Eq.(\ref{reframe})), the on-mass shell condition  for outgoing 
particles (see Eq.(\ref{onmass})), and  conditions of high energy  scattering of Eq.(\ref{highkin}).
 In the large $Q^2$ limit for which 
$k^\prime_{1\, \perp} =  k_{1,\perp} + q_\perp \gg  k_{1,\perp}$ and $k^{\prime 2}_{1, \perp} \approx Q^2$ 
one obtains:
\begin{align}
&\delta (p_N^- + q^- - k^{\prime-}_{1} - \sum_{i=2}^3 k_i^- - k_R^-)  \nonumber \\
&= \delta \bigg( \frac{m_N^2}{p_N^+}- + \frac{2 p_N \cdot q}{p_N^+} - 
\frac{k^{\prime 2}_{1,\perp} +m^2}{x^\prime_{1}p_N^+}  \nonumber \\
&- \sum_{i=2}^3 \frac{k_{i,\perp}^2 +m^2}{x_{i}p_N^+}  -\frac{k_{R,\perp}^2 +m_R^2}{x_{R}p_N^+}\bigg) \nonumber \\
& \approx \delta \left(\frac{2 p_N \cdot q}{p_N^+} - \frac{Q^2}{x_{1}'p_N^++}\right)  \nonumber \\
& = \frac{x_{1}p_N^+}{2 p_N \cdot q} \delta(x_1 - x_B) |_{x_B = \frac{Q^2}{2 p_N \cdot q}}.
 \label{momminus}
\end{align}
Furthermore, introducing the initial transverse momentum of the struck quark 
$\mathbf{k}_{1\perp} = \mathbf{k}^\prime_{1,\perp} - \mathbf{q}_\perp$ for 
the conservation of transverse momentum one obtains: 
\begin{align}
&\delta^{(2)}(\mathbf{p}_{N,\perp} + \mathbf{q}_\perp -  {\bf k}^{\prime}_{1,\perp} - \sum_{i=2}^3 \mathbf{k}_{i, \perp} - \mathbf{k}_{R, \perp}) \nonumber \\ 
&= \delta(\sum_{i=1}^3 \mathbf{k}_{i, \perp} + \mathbf{k}_{R, \perp})
\label{momtrans}
\end{align}
Combining Eqs.(\ref{momplus},\ref{momminus}) and (\ref{momtrans}) in Eq.(\ref{delta4}) and substituting it in Eq.(\ref{Wmunumodel}) one obtains:
\begin{equation}
W_N^{\mu\nu} (x, Q^2) = \frac{1}{4\pi m_N} \sum_{h_i, q} \int  [dx][d^2 \mathbf{k}_\perp]  \delta(x_1 - x_B)
A^{\mu \dagger} A^\nu,
\label{Wmunu2}
\end{equation}
where the integration factors  are defined as: 
\begin{equation}
[dx] =\delta(1- \sum_{i=1}^3 x_i - x_R) \frac{dx_R}{x_R}\prod_{i=1}^{3}  \frac{dx_i }{x_i}  
\end{equation}
\begin{equation}
[d^2 \mathbf{k}_\perp] =  16 \pi^3 \delta^{(2)}(\sum_{i=1}^3 \mathbf{k}_{i,\perp} + \mathbf{k}_{R, \perp})\frac{d^2 \mathbf{k}_{R, \perp}}{16 \pi^3 }\prod_{i=1}^{3}  \frac{d^2 \mathbf{k}_{i,\perp}}{16 \pi^3}. 
\end{equation}

Substituting now the ``+" component of the scattering amplitude from Eq.(\ref{Amu3}) to 
Eq.(\ref{Wmunu2}) and using Eq.(\ref{F2F1}) one obtains for  the $F_2$ structure 
function the following expression:
\begin{align} 
&F_2(x_B) = \sum_q \sum_{h_i} \int [dx] [d^2\mathbf{k}_\perp]  e_q^2 x_1 \delta(x_1 - x_B)\nonumber \\
&\times |\psi_{3q}(\{\beta_i, \mathbf{k}_{i, \perp}, h_i \}_{i=1}^3) |^2 |\psi_V (x_V, \mathbf{k}_{R, \perp}, x_R, \mathbf{k}_{V, \perp})|^2,
\end{align}
which together with Eq.(\ref{partmodel}) allows us  to obtain the expression for the valence PDF in the form:
\begin{align}
&f_q(x_B)= \sum_{h_i} \int [dx] [d^2\mathbf{k}_\perp]   \delta(x_1 - x_B) \nonumber \\
& \times|\psi_{3q}(\{\beta_i, \mathbf{k}_{i, \perp}, h_i \}_{i=1}^3) |^2 |\psi_V (x_V, \mathbf{k}_{R, \perp}, x_R, \mathbf{k}_{V, \perp})|^2.
\label{fq_mf}
\end{align}

\subsection{Modeling the Wave Functions}

\subsubsection{An Approach for  Modeling LF Wave Functions }
\label{wfappro}

There have been many approaches  in modeling  light-front valence quark wave functions  
of hadrons (see e.g. \cite{Lepage:1980fj,Brodsky:1981jv,Pasquini:2008ax}). 
Quantum mechanically, for a bound system the 
expansion of the potential close to the stability point results in a harmonic oscillator~(HO) 
type interaction potential, thus justifying the use of 
the HO eigenfunctions as a basis for modeling nonperturbative wave function of hadrons.  
From a practical point of view,
the advantage of using HO wave functions was that they naturally contain the effect of 
confinement and the Gaussian form was convenient for analytic calculations.
The HO  approach was used both in constituent (e.g. Ref. {\cite{Isgur:1979be})  and current (e.g. Ref.\cite{Lepage:1980fj}) quark  descriptions of hadrons.
For the case of current quark description, one important issue was  a proper relativistic generalization of the wave functions.
For two valence quark systems such as pions, one straightforward step in the relativistic generalization was
introducing  the relative momentum of two valence quarks in the center of mass frame according to 
$k^2 = {m_q^2 + k^2_{\perp}\over x(1-x)} - m_q^2$, which allowed to represent the wave function 
in the boost-invariant form along the  momentum transfer direction 
$\bf q$~\cite{Lepage:1980fj,Brodsky:1981jv}.

In our approach we again use the HO basis for the LF wave functions.
In the relativistic generalization of such wave functions  we introduce the relative momentum similar as it was 
defined above for two-valence quark  system.  However, we also introduce the 
phase space factor for the residual (spectator)  system that allows us to  recover the non-relativistic normalization of the wave function  in the lab 
frame where  the wave function represents the eigenstate of the HO Hamiltonian.  
This phase factor appears also if we relate Weinberg (or Bethe-Salpeter) type equations in the nonrelativistic limit to the   Lippman-Schwinger equation  for bound systems  in the lab frame, where the potential of the interaction is defined  (see e.g. \cite{Sargsian:2001ax}). For example, in considering a generic two-body bound system, the HO based LF wave function is 
represented in the following form:
\begin{equation}
\psi_2(\alpha_1, \mathbf{p}_{1\perp},\alpha_2, \mathbf{p}_{2\perp}) = \sqrt{2(2\pi)^3M_T}  \Psi_0(k_{cm})
 \sqrt{\alpha_2},
 \label{2bodyWF}
\end{equation}
where $\Psi_0$ is the ground state wave function of the non-relativistic HO Hamiltonian in the lab frame of the target with mass $M_T$,  normalized conventionally as:
\begin{equation}
\int |\Psi_0(p)|^2 d^3p = 1.
\label{nonrelnorm}
\end{equation}
In Eq.(\ref{2bodyWF}) the residual (spectator) particle ``2" is treated as on-shell 
with $\alpha_1 = 1-\alpha_2$, ${\bf p}_{1, \perp} = - {\bf p}_{2, \perp}$, and 
$\alpha_2 = {E^{lab}_{2} + p^{lab}_{2z}\over M_T} = {E^{cm}_{2} + k^{cm}_{z} \over E_{2,cm} + E_{1,cm}}$, where 
\begin{equation}
k_{cm}^2 = \frac{(s-(m_1-m_2)^2)(s-(m_1+m_2)^2)}{4s},
\end{equation}  
with 
\begin{equation}
s = (p_1 + p_2)^2 =  \frac{p_{1,\perp}^2 +m^2}{\alpha_1} + \frac{p_{2,\perp}^2 +m^2}{\alpha_2}.\end{equation}
The extra factor of $\sqrt{\alpha_2}$ in Eq.(\ref{2bodyWF}) is to account for the phase space of the residual nucleon that allows to 
relate the relativistic normalization of the LF wave function to the nonrelativistic normalization (\ref{nonrelnorm})  in the lab frame. Namely starting 
with the relativistic normalization:
\begin{equation}
\int |\psi_2(\alpha_1, \mathbf{p}_{1, \perp},\alpha_2, \mathbf{p}_{2, \perp})|^2  \frac{d\alpha_2}{\alpha_2}\frac{ d^2 p_{2,\perp}}{16\pi^3}= 1,
\label{normrel2}
\end{equation}
and using Eq.(\ref{2bodyWF}) one obtains:
\begin{align}
&M_T16\pi^3 \int |\Psi_0(k_{cm})|^2 \alpha_2 \frac{d\alpha_2}{\alpha_2}\frac{ d^2 p_{2,\perp}}{16\pi^3} \nonumber \\ 
& \approx M_T 16\pi^3 \int |\Psi_0(k_{cm})|^2  {d p_{2,z}\over  M_T}  \frac{ d^2 p_{2,\perp}}{16\pi^3}   \nonumber \\
&=  \int  |\Psi(p_2)|^2 d^3p_{2} = 1,
\end{align}
where in the above derivation we used the nonrelativistic limit for  $k_{cm}\to p_2$
and  $\alpha_2 =  \frac{E_2 + p_{2,z}}{M_T}\approx \frac{m_2 + p_{2,z}}{M_T}$ with  
$\rightarrow d\alpha_2 = \frac{dp_{2,z}}{M_T}$.  

It is worth mentioning that the above described prescription 
is not unique to HO wave functions and can be applied to any 
quantum mechanical wave function.

\subsubsection{Wave Function of Three Valence Quark System}
\begin{figure}[ht]
\includegraphics[width= 0.75\columnwidth]{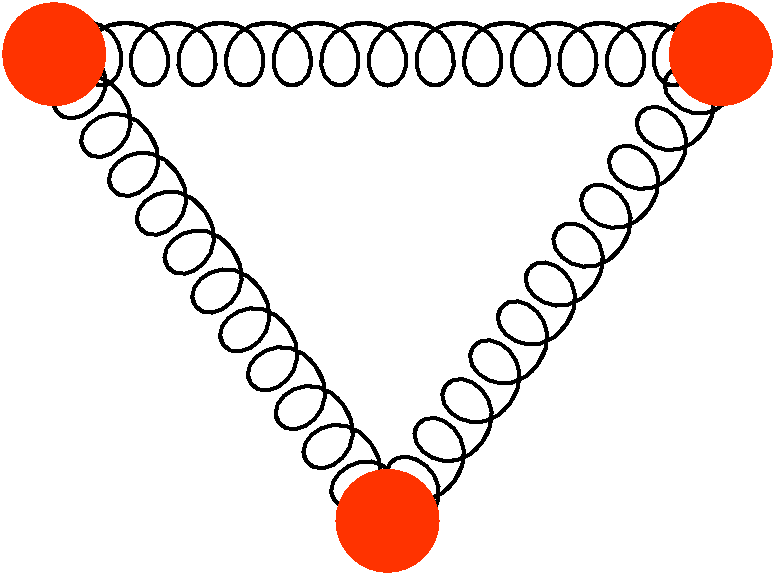}	
\centering
\vspace{-0.2cm}
\caption{Three valence quarks coupled in pairwise HO internation.}
\label{coupledvalencequarks}
\end{figure}

We model the three valence quark system as mutually coupled oscillators 
(Fig.\ref{coupledvalencequarks}) for which the ground state wave function within the above described prescription can be presented in the following form:
\begin{align}
&\psi_{3q} ( \{x_i, \mathbf{k}_{i,\perp}\}_{i=1}^{3}) \nonumber  \\
&= 16\pi^3 m_N A_V \exp{ \left[ -  \frac{B_V}{2}( k_{12,cm}^2 + k_{23,cm}^2 + k_{31,cm}^2) \right]}
\nonumber  \\
&\times \sqrt{x_2 x_3},
\label{3qwf1}
\end{align}
where $A_V$ and $B_V$ are parameters and $x_i, k_{i,\perp}$, ($i\ne j=1,2,3$)  are LC momentum fractions and transverse momenta of each valence quark in the reference frame defined in 
Eq.(\ref{reframe}).
The $k_{ij, cm}^2$s, ($i\ne j=1,2,3$)  in the exponent of the wave function  represent the relative three momenta in the CM system of $i,j$ pairs defined as follows:
\begin{equation}
 k_{ij, cm}^2 = \frac{(s_{ij}-(m_i-m_j)^2) (s_{ij} - (m_i + m_j)^2)}{4s_{ij}},
 \end{equation}
 where the invariant energy of  the $i,j$ pair is:
\begin{align}
s_{ij} &=  (k_i+k_j)^+ (k_i + k_j)^- - (\mathbf{k}_{i,\perp} + \mathbf{k}_{j,\perp})^2  \nonumber\\
&= (x_i + x_j) \left( \frac{k_{i, \perp}^2 + m_i^2}{x_i} + \frac{k_{j, \perp}^2 + m_j^2}{x_j}\right) \nonumber\\
&-(\mathbf{k}_{i,\perp} + \mathbf{k}_{j,\perp})^2.
\end{align}
Since the momentum fractions defined as $x_i = {k^+_{i}\over p^+_N}$, one can introduce total momentum fraction carried  by three-valence quark system as:
\begin{equation}
x_V = x_1 + x_2 + x_3.
\end{equation}
Next we introduce the valence quark transverse momenta defined with respect to the 
total momentum of 3q system $\bf k_V$:
\begin{equation}
 \mathbf{\tilde k}_{i, \perp} = \mathbf{k}_{i, \perp} - \frac{x_i}{x_V} \mathbf{k}_{V, \perp},  \  \ \mbox{(i = 1,2,3)},
 \label{relkperp}
  \end{equation}
  where 
\begin{equation}
\tilde k_{1,\perp} +  \tilde k_{2,\perp} +  \tilde k_{2,\perp} = 0.
\label{kperpsum}
\end{equation}
 Using Eqs.(\ref{relkperp}) and (\ref{kperpsum}) and assuming  same masses for valence quarks one obtains:
 \begin{equation}
s_{12} + s_{23} + s_{31} = \sum_{i=1}^3 x_V\frac{\tilde k_{i, \perp}^2 + m^2}{x_i} + 3m^2,
\end{equation}
and
\begin{equation}
k_{12,cm}^2 + k_{23,cm}^2 + k_{31,cm}^2= \frac{1}{4} \big( \sum_{i=1}^3 x_V\frac{\tilde k_{i, \perp}^2 + m^2}{x_i} - 9m^2 \big).
\end{equation}
Inserting the above relation into Eq.(\ref{3qwf1})  the wave function of 3q- state can be expressed as follows:
\begin{align}
&\psi_{3q} ( \{ x_i, \mathbf{k}_{i, \perp} \}_{i=1}^3) \nonumber \\
&=  16\pi^3 m_N A_V \exp \left[-\frac{B_V}{8} 
\left( \sum_{i=1}^3 x_V\frac{\tilde k_{i, \perp}^2 + m^2}{x_i} - 9m^2 \right)   \right] \nonumber \\
&\times \sqrt{x_2 x_3}.
\label{3qwf2}
\end{align}

\subsubsection{Wave Function of Recoil System}
\label{VRwavefunction}

Using the same approach (Sec.\ref{wfappro}) we model the wave function of the recoil system in the following form:
\begin{equation}
\psi_R (x_R, \mathbf{p}_{R, \perp}) = \sqrt{16\pi^3 m_N} A_R e^{- \frac{B_R}{2} p_R^2} \sqrt{x_R} 
\label{recoilWF}
\end{equation}
where $x_R$ is the light-cone momentum fraction of the nucleon carried by the recoil system and 
$p_R^2 = k_{R, \perp}^2 + p_{R, z}^2$ is its three momentum in the CM of the nucleon. 
As a first approximation that allows to perform calculations analytically,  
we use a non-relativistic kinematics for the recoil system.
%we treat the system  non-relativistically. 
In this case  
in the CM frame of the nucleon one can approximate $x_R \approx {m_R + p_{R,z}\over m_N}$ and substitute
\begin{equation}
p_{R,z} = (x_{R}m_N- m_R).
\label{nonrel}
\end{equation}
In numerical evaluations we will treat $B_R$ and $m_R$ as  free parameters characterizing the recoil system and evaluate $A_R$ from 
the normalization condition similar to Eq.(\ref{normrel2}).

\subsection{Completing the Calculation of Valence PDF}
Substituting the wave functions of 3q (Eq.(\ref{3qwf2})) and the recoil (Eq.(\ref{recoilWF}))  systems in Eq.(\ref{fq_mf}) for the valence quark 
partonic distribution one obtains:
\begin{align}
&f_q(x_B)= \left( 16 \pi^3m_N\right)^3
 |A|^2  \int^{Q^2} 16 \pi^3\delta(x_1-x_B)\nonumber \\
&\delta(1-\sum\limits_{1}^3x_i - x_R)\delta^2(\sum\limits_{1}^3k_{i,\perp}+k_{R,\perp})  \nonumber \\
&\times  
\exp \bigg[-{B_V\over 4}\sum_{i=1}^3 x_V \frac{\tilde k_{i,\perp}^2 +m_i^2}{x_i} \nonumber \\
&- B_R \left( (m_N x_R - m_R)^2  + k_{R,\perp}^2 \right)     \bigg] 
x_2x_3x_R\nonumber \\
& \times \prod\limits_{i=1}^3 {dx_i\over x_i} {d^2k_{i,\perp}\over 16 \pi^3}{dx_R\over x_R} {d^2k_{R,\perp}\over 16 \pi^3},
\label{fqinp1}
\end{align}
where $A = A_R A_V e^{{9\over 4} B_Vm_q^2}$. Note that the transverse momenta ${\bf k}_{i,\perp}$ and ${\bf \tilde k}_{i,\perp}$ are defined  in the CM of nucleon and 3q- system respectively and related 
to each other according to Eq.(\ref{relkperp}). 
 Using Eq.(\ref{relkperp})  and relation ${\bf k}_{V,\perp}   =  - {\bf k}_{R,\perp}$  and $d^2 \mathbf{k}_{i, \perp} = d^2 \mathbf{\tilde k}_{i, \perp}$
 allows  us to factorize  integrations by  $\tilde k_{i,\perp}$ and $k_{R,\perp}$ resulting in:
\begin{align}
 &  f_q(x_B, Q^2)  =  m_N^3 |A|^2 \int\prod\limits_{i=1}^3 {dx_i\over x_i} \delta(x_B-x_1) \nonumber \\
&\times \exp \left[ -\frac{B_V}{4} \sum_{i=1}^3  x_V\frac{ m_i^2}{x_i} - B_R m_N^2(x_R - \frac{m_R}{m_N})^2 \right]  \nonumber \\
&\times x_2 x_3  \int^{Q^2} \prod\limits_{i=1}^3 d^2\tilde k_{i,\perp} \delta^2(\sum\limits_{i=1}^3 {\bf \tilde k}_{i,\perp} )\nonumber \\
&\times \exp{ \left[ -\frac{B_V}{4} \sum_{i=1}^3 x_V \frac{\tilde k_{i,\perp}^2}{x_i}\right]}
\int^{Q^2}  d^2k_{R,\perp} \exp{\left[ - B_R k_{R,\perp}^2  \right]}. 
\label{f2ev2}
\end{align}
Furthermore we evaluate the following integrals in the above equation (see Appendix B):
\begin{align}
&   \int^{Q^2_R} \exp{\left[- B_R k_{R,\perp}^2\right]} d^2p_{R,\perp}   =  {\pi\over B_R}(1-e^{-B_R Q^2_{R,Max}});\nonumber \\
& \int^{Q^2} \prod\limits_{i=1}^3 d^2\tilde k_{i,\perp} 
\delta^2(\sum\limits_{i=1}^3 {\bf \tilde k}_{i,\perp} )
 \exp{ \left[ -\frac{B_V}{4} \sum_{i=1}^3 x_V \frac{\tilde k_{i,\perp}^2}{x_i}\right]} \nonumber\\
& =  {16\pi^2\over B_V^2} {x_1x_2x_3\over x_V^3}\nonumber \\ 
&\times(1-e^{-a_{cm}Q^2_{cm,Max}})(1-e^{-a_{rel}Q^2_{rel,Max}}),
\label{perpints}
\end{align}
where $a_{cm} = {B_V x_V\over 4} {x_V\over x_3(x_1+x_2)}$  and $a_{rel} = {B_V x_V\over 4} {x_1+x_2\over x_1x_2}$.
Here $Q^2_{R,Max}$ is the maximal transverse momentum square of the recoil system. Also in the  derivation 
(see Appendix B) we defined 
${\bf \tilde k}_{12,\perp}^{cm} = {\bf \tilde k}_{1,\perp} + {\bf \tilde k}_{2,\perp}$ and 
${\bf \tilde k}_{12,\perp}^{rel} = 
{x_2{\bf \tilde k}_{1,\perp} - x_1 {\bf \tilde k}_{2,\perp}\over x_1 + x_2}$ and $Q^2_{cm,Max}$ and $Q^2_{rel,Max}$ 
represent the maximal transverse momentum squares of the center of mass and relative transverse momenta of ``1" and ``2" quark system.
Inserting the expressions of Eq.(\ref{perpints}) into Eq.(\ref{f2ev2}) and taking the integral over  $d x_1$ using 
$\delta(x_B-x_1)$ (see Appendix~B for details) one arrives at:
\begin{align}\label{PDF-with-finite-Q2}
& f_q(x_B, Q^2) = \mathcal{N} \displaystyle\int_0^{1-x_B}dx_2 \displaystyle \int_0^{1-x_B - x_2} dx_3 \nonumber \\
&\times \exp \left[ -\frac{B_Vx_V}{4} \sum_{i=1}^3 \frac{ m_i^2}{x_i} - B_R M_N^2(x_V - (1- \frac{M_R}{M_N}))^2 \right] \nonumber \\
&\times  \frac{x_2 x_3}{x_V^3} \left(1-e^{-a_{cm}Q_{cm}^{max 2}}\right)\left(1-e^{-a_{rel}Q_{rel}^{max 2}}\right)\nonumber \\
&\times \left(1-e^{-B_R Q^2}\right),
\end{align}
where $x_1= x_B$ and $x_V= x_B + x_2 + x_3$ and  the normalization constant 
\begin{equation}
\mathcal{N} = {16\pi^3 A_V^2 A_R^2 m_N^3\over  B_R B_V^2} e^{{9\over 4} B_V m_q^2}.
\label{Norma}
\end{equation}
The above equation simplifies further when $Q^2$ is large so that terms with it in the exponential are negligible in Eq.\ref{PDF-with-finite-Q2}:
\begin{align} \label{f-Q2inf}
&  f_q(x_B, Q^2)  =   \mathcal{N} \int_0^{1-x_B}dx_2 \int_0^{1-x_B - x_2} dx_3  \frac{x_2 x_3}{x_V^3}\nonumber \\
&\times \exp \left[ -\frac{B_Vx_V}{4} \sum_{i=1}^3 \frac{ m_i^2}{x_i} - B_R M_N^2(x_V - (1- \frac{m_R}{m_N}))^2 \right]  
\end{align}
We further simplify this expression  by considering the massless limit of valence quarks and changing the integration  variable from $x_3$ to $x_V$. 
This allows us to evaluate the $dx_2$ integration analytically  resulting in the final expression for the valence quark distribution in the following form:
\begin{align}
& f_q(x_B, Q^2) = {{\cal N}\over 6} \int\limits_{x_B}^1 dx_V {(x_V-x_B)^3\over x_V^3}\nonumber \\
& \times  \exp\left[ -B_R m_N^2\left(x_V - (1- \frac{m_R}{m_N})\right)^2 \right]. 
\label{f_qfin}
 \end{align}
Note that  as it follows from the above equation, in  the massless limit, 
the parameter $B_V$ enters in the structure function only through the normalization factor 
$\cal N$.

 \subsection{Qualitative Features of the Model}

One important  feature  of the model can be observed if we consider Eq.(\ref{f_qfin}) 
at $x_B\sim x_p$, when $x_V\sim 1$. In this case   the qualitative behavior of  $f_q(x_B, Q^2)$  can be obtained by
evaluating Eq.(\ref{f_qfin}) at  the maximum of the exponent;
$x_V = 1- \frac{m_R}{m_N}$. This results in $f_q(x_B,Q^2)\sim (1-  x_B - {m_R\over m_N})^3$ and for 
the valence quark structure function one obtains:
\begin{equation}
h(x_b,t) = x_B f_q(x_B,Q^2)\sim x_B\left(1-  x_B - {m_R\over m_N}\right)^3.
 \end{equation}
From the above relation it follows that $h(x,t)$ function peaks at
\begin{equation}
x_{p} \approx {1\over 4}(1-{m_R\over m_N}).
\label{peakpion}
\end{equation}
Thus one observes that within the considered model the peak of the $h(x,t)$ distribution for 
valence quarks  (see Fig.\ref{peak_and_height}(b))  is related to the mass of the residual system in the nucleon.

Now if we use the characteristic value for the peak position $x_{p} \simeq 0.2$ for valence quarks at  starting $Q^2=4$~GeV$^2$, from Eq.(\ref{peakpion})  one observes that it will correspond to $m_R\approx m_\pi$. 
Thus the present model mimics a pion-cloud like picture for the residual system.

It is interesting that in Ref. \cite{Close:1988br} it was observed that within diquark model 
the magnitude of $x_p$ is related to the mass of the recoil diquark in the form:
\begin{equation}
x_p \approx (1- {m_d\over m_N}),
\label{diquark}
\end{equation}
 where $m_d$ is the mass of the diquark.  This and Eq.(\ref{diquark}) however are different,  reflecting 
 different dynamics of valence quark generation in the nucleon.

\medskip

It is worth mentioning that with an increase of $Q^2$, because of   the radiation of valence quarks the magnitude of $x_V$ diminishes and due to the  ${(x_V-x_B)^3\over x_V^3}$ factor in 
Eq.(\ref{f_qfin}) the $x_p~-~m_R$ relation become more complicated than 
Eq.(\ref{peakpion}).
\medskip
\medskip

Next we evaluate valence quark PDFs at $x_B\rightarrow 1$ limit.  For this we substitute  
$x_B = 1 - \epsilon$ in Eq.(\ref{f_qfin}) and in the $\epsilon \rightarrow 0$ limit evaluate the 
integral  which results in 
\begin{equation}
f_q(x_B, Q^2)\mid_{x_B\rightarrow 1} = {{\cal N}\over 24}e^{ -B_R m_R^2} (1-x_B)^4.
\label{xto1}
\end{equation}
This result indicates that the considered mean field model predicts a fall off of the valence quark 
PDF faster than the one predicted within perturbative QCD\cite{Lepage:1980fj};  $(1-x_B)^3$.
Thus the observation of the  $(1-x_B)^4$ behavior at $x_B\rightarrow 1$ limit will indicate 
the dominance of the mean field dynamics.

\section{Numerical Estimates of Valence quark distributions}
\label{estimates}

\subsection{Strategy of choosing the parameters of the model}
\label{paramstategy}

We use Eq.(\ref{f_qfin}) as a baseline formula for  fitting to the phenomenological PDFs evaluated in leading order.
%,  for which we choose  CT14 parametrization\cite{Dulat:2015mca}. 
In general the model has 
five parameters $A_V$ $B_V$, $A_R$, $B_R$ and $m_R$.

For the valence quarks we assume that characteristic separations in the 3q system in the impact parameter space 
is $\langle b^2_{i,j}\rangle \sim$~(0.3Fm)$^2$.  This allows us, based on 
 Eq.(\ref{3qwf2}), to  evaluate $B_V = 4 \langle b^2_{i,j}\rangle {x_i\over x_V} \approx 
{4\over 3} \langle b^2_{i,j}\rangle \approx 3.08$~GeV$^{-2}$.  

For the recoil system we  relate  $A_R = \left({B_R\over \pi}\right)^{3\over 4}$.  
%As we observed in the previous section, the parameter characterizing the mass of the residual system, $m_R$ is related to 
% the position of the peak of the $h(x,t)$ function.
Finally, the remaining parameter $A_V$ is fixed through the normalization factor, ${\cal N}$ and Eq.(\ref{Norma}), yielding:
\begin{equation}
A_V =\sqrt{B_R{\cal N}\over  16 \pi ^3 m_N ^3} {B_V\over A_R}
\label{AV}
\end{equation}

In the fitting procedure the parameters ${\cal N}$, $m_R$ and $B_R$ are evaluated by fitting  Eq.(\ref{f_qfin}}) to the height, position 
and the width of the $h(x,t)$ distribution in leading order approximation.

\begin{figure*}[ht]	
\includegraphics[width=1.0\textwidth]{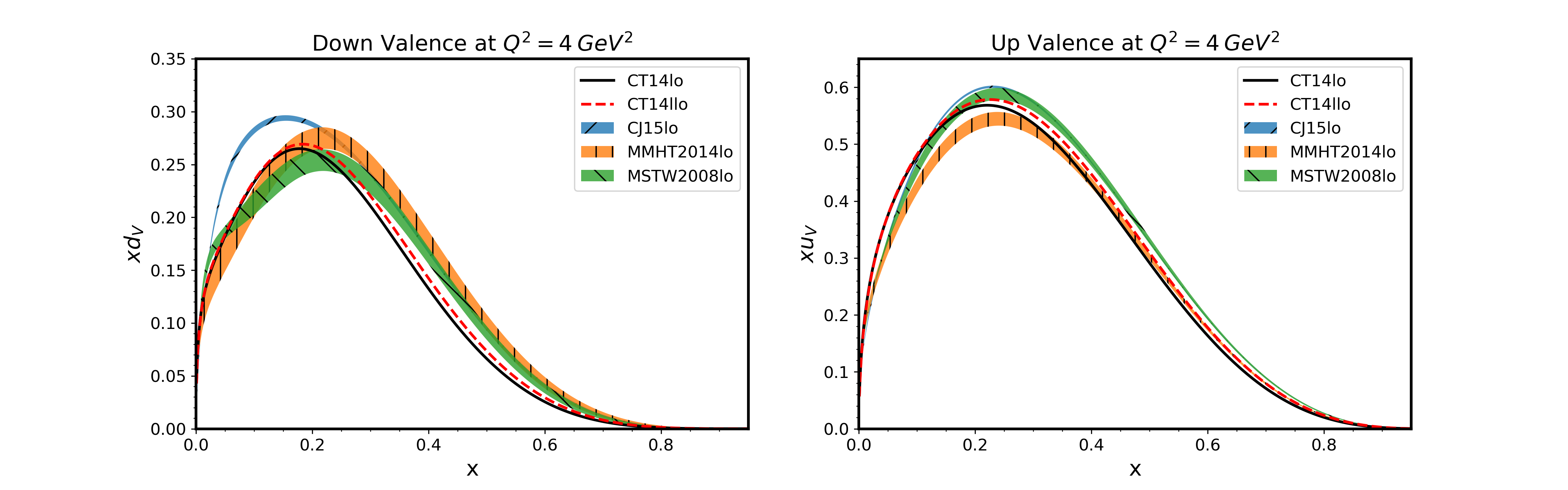}
\centering
\caption{The h(x) distribution at $Q^2=4$~GeV$^2$. Left panel for valence d- quark distribution. Right panel 
for valence u-quark distributions.}
\label{h_lo_dists}
\end{figure*}

As a starting $Q^2$ we choose $Q^2=4$~GeV$^2$ such that it provides large enough produced mass 
$W\ge 2$~GeV in the range of  up to $x \sim 0.65$ for which  we expect that mean-field dynamics have 
non-neggligible contribution. This $Q^2$ is also large enough that it resolves the $3q$ valence cluster in the nucleon,
for which one requires $Q^2\gg {1\over B_V}$.
In Fig.\ref{h_lo_dists} we show the current status of the valence PDFs in LO for d- and u- quarks evaluated by using
modern PDF parameterizations\cite{Accardi:2016qay,Martin:2009iq,Dulat:2015mca,Owens:2012bv,Harland-Lang:2014zoa} 
at $Q^2=4$~GeV$^2$. As the figure shows the  height and the peak-position of the distributions 
vary between different parameterization  on the level of 15-20\%. Thus by fitting to these distributions we 
evaluate the range of parameters as they change from one to other parameterizations.
We fit using maximum likelihood  approximation without  estimating the  errors of parameters for specific
PDF parameterization.  In such way our estimates can be considered qualitative (see also discussion in Sec.V).

%calculated using CT14 leading order  parameterization\cite{Dulat:2015mca}.

\subsection{Fitting results and estimation of  the magnitude of high momentum component of the valence quark distributions}

The results for parameters $\cal{N}$, $B_R$ and $m_R$ for considered PDF parameterizations are given in 
Table~\ref{table1}.  As these parameters show despite their variations, there
are common features shared by different parameterizations.  The one is that the residual mass 
for the case of valence $u$-quarks  ($0.04 \le m_R^u \le 0.07$~GeV) is less than the corresponding mass for 
the $d$-quark ($0.17 \le m_R^d \le 0.32$~GeV).  Additionally, $m_R^u\le m_{pion}$ and $m_R^d > m_{pion}$.
These evaluations  predict that in the case of scattering from valence d- quark in the  proton 
there is a larger probability to detect recoil  pions than in the case of scattering from the u- quark. 

\begin{table*}%[ht]
\caption{Fitting parameters for valence d- and u- quarks.} 
\centering 
\begin{tabular}{l c c c c c c c} % centered columns (4 columns)
\hline\hline %inserts double horizontal lines
 d-quark  \ &  N$^d$ & \  B$_R^d$(GeV$^{-2}$) & \  m$_R^d$ (GeV)  & \ u-quark &  \  N$^u$ & \  B$_R^u$ (GeV$^{-2}$) & \ m$_R^u$ (GeV)  \\% [0.5ex] % inserts table
\hline\hline %inserts double horizontal lines
 CT14LL0\ & 64  & \  30  & \  0.26 &   \    & 174 & \ 42 & \ 0.07 \\ 
\hline % inserts single horizontal line
CT14L0  \ & 63 &   \  29  & \  0.24  &  \    &  183  & 42 & \ 0.06   \\ %[0.5ex] 
\hline % inserts single horizontal line
CJ15lo            \ & 47   &  \  6    & \  0.32  &   \    & 208 & \ 50 &\ 0.05 \\ 
\hline % inserts single horizontal line
 MMHT2014lo\  & 76   &  \  50  & \  0.16  &   \    & 228 & \ 60 & \ 0.04 \\ 
\hline % inserts single horizontal line
MSTW2008lo\   & 71   &  \  50  & \  0.17  &   \    & 235 & \ 65 & \ 0.04 \\ 
\hline % inserts single horizontal line
\hline %inserts single line
\end{tabular}
\label{table1} % is used to refer this table in the text
\end{table*}

Another characteristics of the parameters is that the slope factor for the residual system  in the  case of  scattering 
from  valence u-quarks ($42 \le  B_R^u \le 65$~GeV$^{-2}$) is systematically larger than for  the case of 
d-quarks ($29 \le  B_R^d \le 50$~GeV$^{-2}$) 
(not included  is the CJ15 parametrization  for the d-quarks which  yields  $B_R^d \le 6$~GeV$^{-2}$.).
This indicates a more compact state for the case of scattering from d quarks which can be checked by 
measuring the attenuation of the residual system in the nucleus depending whether scattering took place on valence 
d- or u- quark.

\begin{figure}[ht]	
\vspace{1.0cm}
\hspace{-1.0cm}
\includegraphics[scale=0.42]{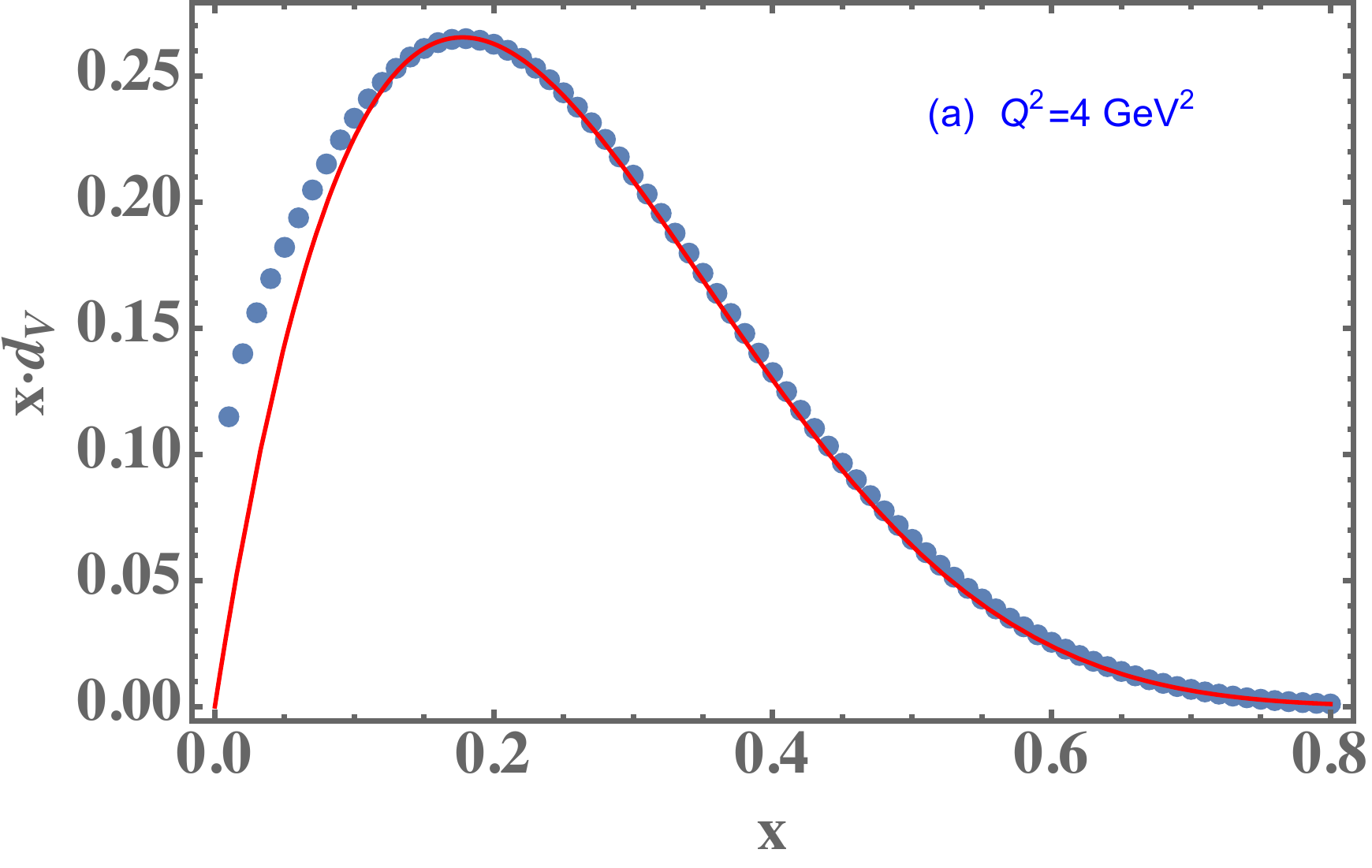}
\vspace{-2cm}
\hspace{1cm}
\includegraphics[scale=0.4]{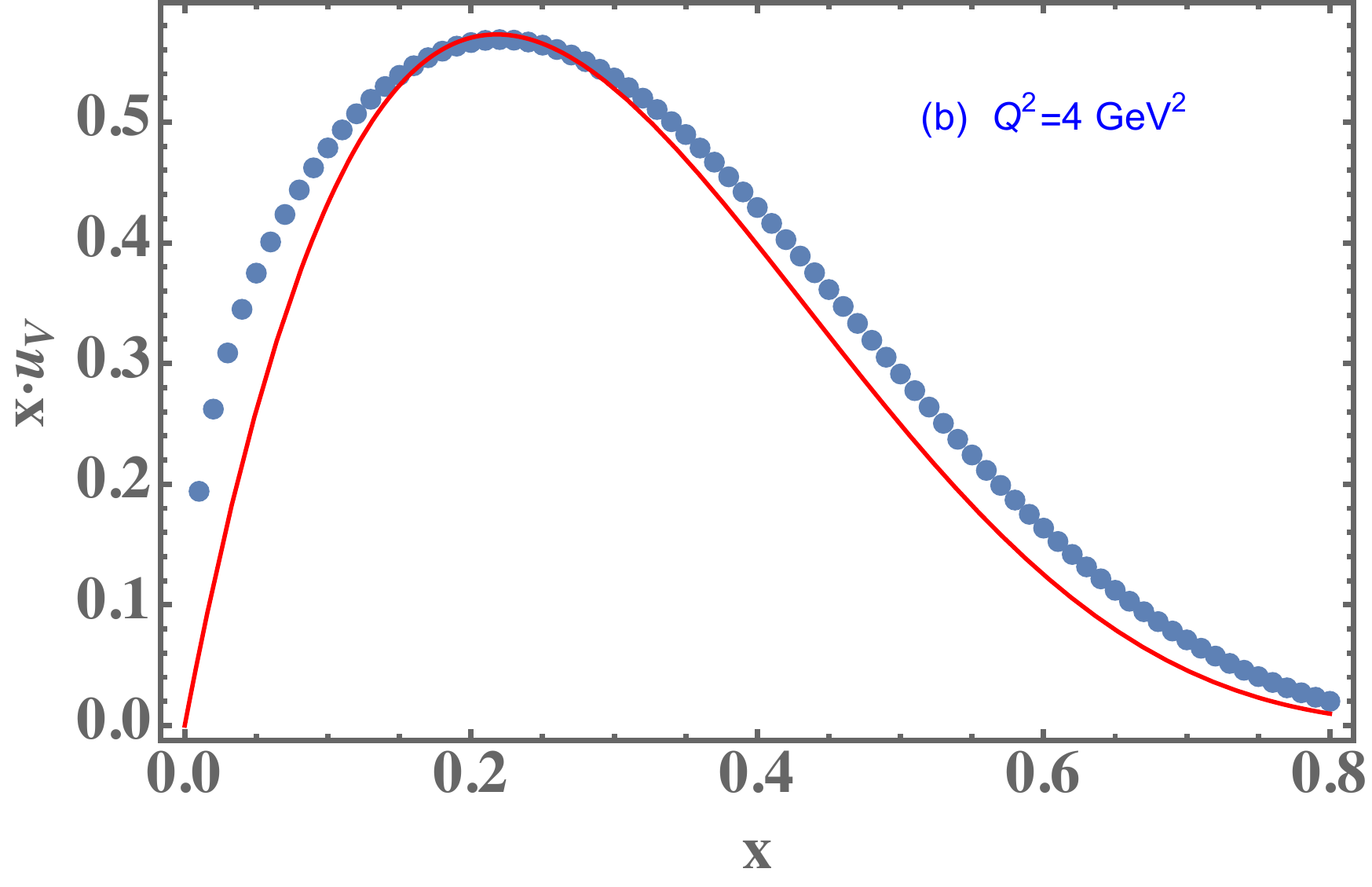}
\vspace{1.4cm}
\centering
\caption{Comparison of the results of the fitting of mean-field model to  the $x q_V(x)$ distribution for d- (a) and  u- (b) quarks. Data points corresponds to the CT14nnlo parameterization for central PDF sets and the shaded areas shows the uncertainties of 
PDFs estimated through the PDF error sets (for details see Ref.\cite{Dulat:2015mca}).}
\label{xvfit}
\end{figure}

In Fig.\ref{xvfit}  we present a comparison of the  model with the CT14nnlo PDFs at $Q^2=4$~GeV$^2$\footnote{Similar 
picture is obtained for other parameterizations, except for the valence $d$-quark distribution in the case of CJ15 which 
shows  wider distribution with lower position of the peak,}
As the figure shows,  a better description is achieved for the valence d-quark than for the u-quark distribution. 
This indicates that larger  high momentum component is needed for valence u-quarks  than for the d-quark.

To evaluate the expected contribution from the high momentum component of valence quark distribution, 
we used the  fitting parameters to  evaluate normalization   and momentum sum rules for both valence d- and u- quarks.  
For the  normalization  one obtains  for the d-quark $N_d = 0.64$ and for the u-quark $N_u = 1.37$.  
It is interesting that these normalizations result in a 
${N_d\over N_u}\sim 0.5$.  These results also indicates that one expects that Regge mechanism at $x<x_p$ 
together with   $qq$-correlations 
to contribute  $\sim$ 36\% and $\sim$32\%  of total normalizations for d- and u- quarks respectively.

The evaluation of the momentum sum rule ($P_q = \int x q_V(x) dx$)  yields $P_{d} = 0.095$ and $P_u = 0.24$ which 
should be compared with $P_{d} = 0.1$ and $P_{u}= 0.268$ obtained from the CT14llo distribution.  
Expecting that the Regge mechanism, which dominates at 
$x<x_p$, has a smaller contribution to the momentum sum rule the above estimates 
allow us to evaluate better  the contribution from $qq$ correlations.  As the  mentioned numbers show, 
the expected contribution from   $qq$ correlations in the 
momentum sum rule  for the  d-quark is   $\sim$~5\%  and for the u-quark is $\sim$~10\%.   
These evaluations however can be somewhat of an underestimation, since in fitting the model to the PDFs we assumed that all the 
strength of PDF at $x=x_p$ is due to the mean-field mechanism. The larger contribution for quark- correlation mechanism may require a renormalization of  the mean field distribution. 
This problem can be addressed after modeling high-x tail of valence PDFs and  combining it with the  current mean-field model.
Overall our current result indicates an expected  larger contribution of high momentum tail  to the valence u- quark distribution 
compared to that of d-quarks.

\medskip

Another interesting observable is the ratio of the d- to u- quarks at $x\to 1$ limit. As it follows from Eq.(\ref{xto1}) the model 
predicts
\begin{equation}
{f_d(x)\over f_u(x)} \mid_{x_B\rightarrow 1}  = {N_d\over N_u}e^{\left (B_R^u(m_R^u)^2 -B_R^d (m_R^d)^2\right)}.
\end{equation}
The interesting feature of the model is that  the properties of valence PDFs (the mass of the residual system, the slope factor and normalization constant) at the vicinity of the peak define the magnitude of the $d/u$ ratio at x=1.
The mechanism that results in 
the decrease of  ${d_V\over u_V}$ ratio with an increase of x, is  due to the inequality of Eq.(\ref{udmass}) because of which 
the recoil wave function for the case of  the d- quark decreases faster than that of  the u- quark.  Thus the value of the 
${d_V\over u_V}$ ratio at $x\to 1$   in the present model is  related to the difference in the peak positions for $h(x,t)$ distribution for 
d- and u- quarks.

Using parameters of Table~\ref{table1} one evaluates
\begin{equation}
0.06 \le {d_V\over u_V}\mid_{x\to 1} \le 0.1 (0.14)
\end{equation}
where $(0.14)$ is the estimate for CJ15 parameterization. 
This is the prediction of the current mean-field model if 
quark-quark correlations will be found to be negligible at large $x$.

\begin{figure}[ht]	
\includegraphics[width=\columnwidth]{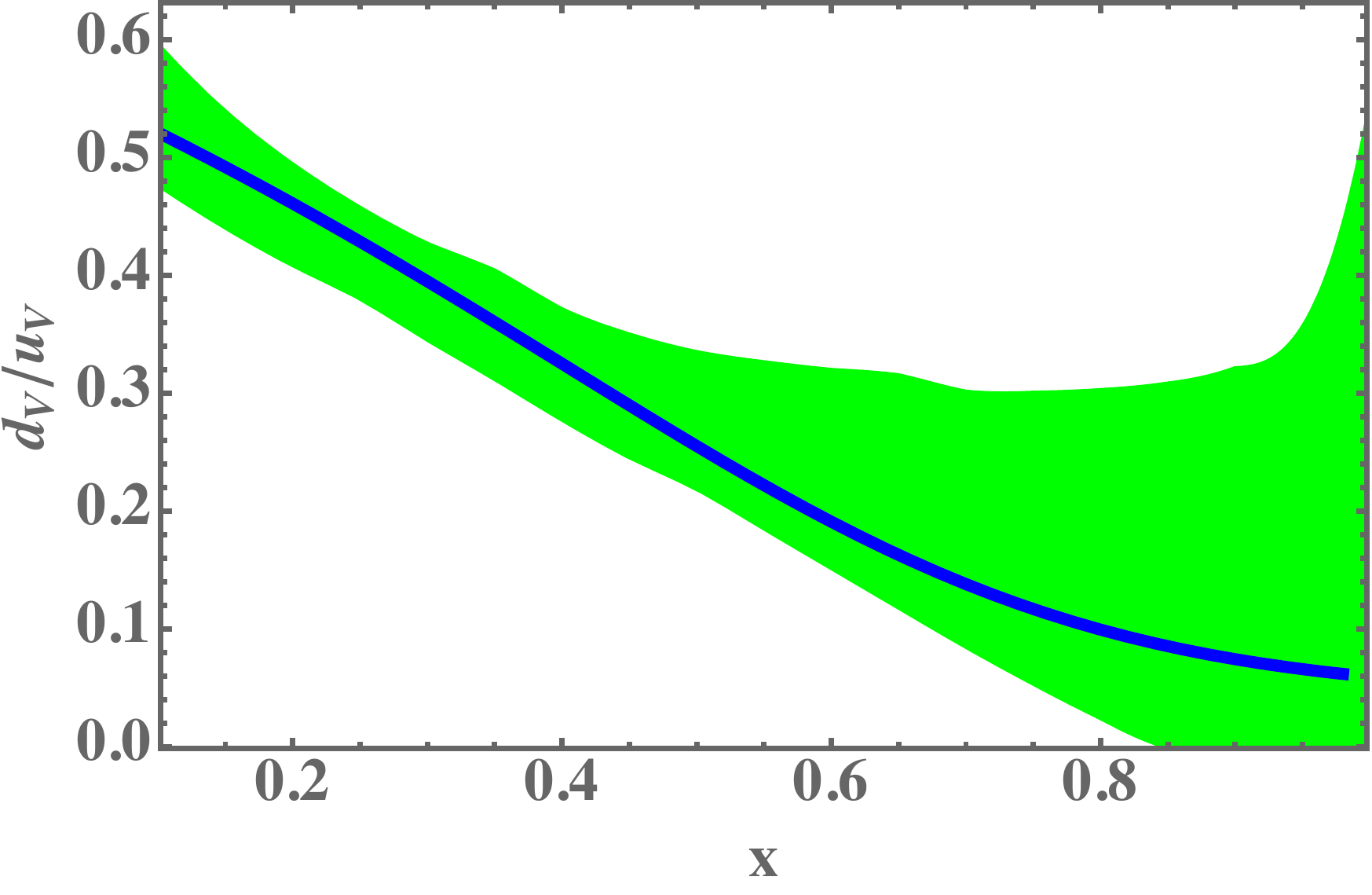}
\vspace{-0.4cm}
\caption{The $x$ dependence of ${d_V\over u_V}$ ratio in valence quark region, solid curve. The calculation is compared with 
the one following from CT14nnlo  parametrization. The  area takes into account the  spread of the PDFs for d- and u- quark distributions\cite{Dulat:2015mca}.}
\label{doveru}
\end{figure}

In Fig.\ref{doveru} we present the $x$ dependence of the ${d_V\over u_V}$ ratio for the same parameters used in 
Fig.\ref{xvfit}.
If, however, the current estimates of $d/u$ ratio  from DIS processes off $A=3$ nuclei
\cite{Abrams:2021xum,Segarra:2021exb}  is  confirmed, 
it will indicate the need of  $qq$ short range interactions to enhance the $d/u$  ratio at $x=1$, 
to the values, $\sim 0.2$.

 \section{Summary and Outlook}
\label{outlook}

We have developed a model which describes the mean-field dynamics of  valence quarks in the nucleon by
separating the nucleon into a valence 3q- cluster   and residual system.  
Within this model based on effective light-front diagrammatic approach we calculated the valence quark distributions 
in the nucleon  in leading order approximation. The parameters entering the model are ones that 
describe  the wave functions of nonperturbative 3q and residual systems.
\subsection{Summary of results and predictions of the model}
An  important result  of the model is the  prediction  of  a new relation for the position of the peak 
in the $x q_V(x)$ distribution and the mass of the residual system according to Eq.(\ref{peakpion}) for moderately 
large $Q^2$.
This relation naturally describes  the fact that the valence d- quarks peak at lower $x_p$ than u- quarks following from the expectations,  that in the proton,  the  residual system
for the d-quark in general has  larger mass than for the  u-quarks (Eq.(\ref{massspectrum})). 

The model predicts $(1-x_B)^4$ behavior for valence quark PDFs at $x_B\rightarrow 1$,
which is stronger than one predicted in pQCD, $(1-x_B)^3$. This indicates 
the  need of  inclusion of 2q and 3q correlations to describe high x tail of valence PDFs.

 To evaluate the parameters of the model we fitted the calculated valence quark distributions to the 
 several leading order phenomenological PDFs with starting $Q^2=4$~GeV$^2$.  The obtained parameters 
 depend on the particular choice of the phenomenological PDF. However they demonstrate the 
 following common features 
 \begin{itemize}
 \item The mass of the residual system for probed valence u- quarks is systematically lower than that of 
 the valence d-quarks and it is less than the pion mass
 \item The slope factor of the residual system  in the proton is larger for the u-quarks than for the d-  quark
  indicating that d-quark- residual system is more compact
 \item The magnitude of the parameters indicate the need for larger (by factor of two) high momentum component 
for the valence u-quarks than for the d-quark.  
 \end{itemize}
 The above features allow us to make specific predictions that can be checked experimentally. One is that 
 there should be a correlation between the multiplicity of pion productions in the target fragmentation region with 
 the flavor of the knock-out quark in the forward direction. Specifically one expects more pions to be produced 
 in the target fragmentation region of the proton  if the forward scattering takes place on the valence 
 d-quark.  Another prediction  of the model is that  for asymmetric nuclei, attenuation of the residual system is correlated 
 with the flavor of struck quark in the forward direction.  This prediction is based on the observation that the 
 recoil system is more compact  (larger $B_R$)  in the case of interaction with the valence d-quark in the 
 proton.

\subsection{Possible  improvements and fitting  to beyond the leading order PDFs} 
To be able to calculate the nucleon structure function analytically  we treated the kinematics of 
recoil system non-relativistically (Eq. (\ref{nonrel})). However the fitting results of $m_R$ indicate the 
need of relativistic treatment of the recoil system. This will be done in the process of fitting the model 
parameters to the  next to leading order PDFs.  For the latter case since all modern NLO PDFs adopt 
$\overline{MS}$ factorization scheme the relation between structure function $F_2$ and PDF, $f_i(x,t)$ is a convolution integral
which covers region above the  probed Bjorken x kinematics. Thus, fitting to the NLO PDFs will require the modeling of the 
contribution from $qq$ correlations presented in Fig.\ref{framework}(b) and (c).  These calculations are in progress 
and will be published in follow-up paper.

 Finally, the reliability of the obtained parameters  for the 3q-cluster and residual system wave functions can be checked 
 by calculating the mean-field effects at $x\sim 0.1-0.4$, in many different quantities in QCD, such as form-factors, generalized 
 partonic distribution functions or transverse momentum distributions for different DIS processes.   
 
With the addition of the two- and  three- quark  short-range interactions to the present model we hope to 
have a framework that will allow us to study the dynamics of the valence quarks in the all 
range of $x>0.1$. The  parameters obtained in this work, for the wave functions of  mean field  3q-cluster (Eqs. (\ref{3qwf2})) 
and residual system~(Eq. (\ref{recoilWF}))  will allow us to use them in the calculation of the correlation diagrams of 
Fig.(\ref{framework})(b) and (c). Note that in this work we did not use polarized PDF's to fix the parameters of the model 
since one of the assumptions was that the mean-field dynamics adheres SU(6) symmetry.  However once two- and three- quark 
short-range interactions are added they will break the SU(6) symmetry because of vector exchange nature of 
interactions. Then by fitting  calculated distributions to  polarized PDFs we will be able to calibrate 
contribution from short range interaction at large $x$.

%\acknowledgments

\begin{acknowledgements}
This work is supported by United States  Department of Energy  grant under contract DE-FG02-01ER41172.
\end{acknowledgements}

%This work is supported by United States  Department of Energy  grant under contract DE-FG02-01ER41172.

%\section*{Appendix}
\appendix

\section{Light-Front Diagrammatic Rules}

The light-front diagrammatic  rules are given  for the processes ordered in light-cone time, 
$\tau = x^+ = t + z$ with the possibility of an instantaneous interaction.

\begin{enumerate}
\item  Draw  topologically distinct $\tau$-ordered diagrams identifying effective interaction vertices.
In addition to the usual advanced and retarded propagation between two events one needs to include a  possibility in which the two events connected by an internal fermion or photon 
lines that interact at the same LC $\tau-$time, commonly referred as an instantaneous term.
\item For each particle line assign four-momentum $p^\mu$ and spin $s$ (or helicity, $\lambda$) 
with the on-mass-shell conditions $p^2=m^2$.
\item Incoming fermions/anti-fermions get $u(p,h) / \bar{v}(p,h)$ . Outgoing fermions/anti-fermions have 
$\bar{u}(p,h) / v(p,h)$. 
\item Incoming photons or gluons with polarization $\lambda$ are assigned with polarization 
vector $\epsilon^\mu_\lambda(k)$
\item For each internal particle line assign a factor  $\frac{\theta(p^+)}{p^+}$
\item Intermediate states get a light front energy denominator of:
$${1\over \mathcal{D}} =  \frac{1}{\sum_{init.}p^- - \sum_{interm.}p^- + i\epsilon}$$
where sums go over the initial and intermediate states.
\item  Assign to the each  vertex in the diagram the effective transition factor $\Gamma$. 
For bare particle elementary interactions these factors will correspond to the 
fundamental vertices with coupling constants.
\item For the transition of the composite particle, A with momentum $p_A$ and helicity $h_A$ to  its n-constituents one introduces  LF wave function in the form:
\begin{equation}
 \psi(\{ x_i,k_{i,\perp},h_i\}_i^n) = 
{\prod\limits_{i=1}^{n} \chi_{fi}(x_i,k_{i,\perp},h_i)\Gamma \chi_A(p_A,h_A) \over p_{A}^+  \mathcal{D}},
\end{equation}
where $x_i$ and $k_{i,\perp}$ are LC momentum fractions and transverse momenta of 
constituent particles with helicity $h_i$. The spin wave functions of outgoing particles 
are described by  $\chi_{fi}(x_i,k_{i,\perp},h_i)$ and $\Gamma$ is the effective vertex of the 
transition of particle A to n-constituents.
\item Each closed loop is integrated by the LC space factor
$$\int \frac{dp^+d^2 \mathbf{p_\perp} }{16\pi^3} $$
\end{enumerate}

\section{Integrations in the transverse momentum space
}
Now, if we introduce the relative  transverse momentum 
$\mathbf{k}_{i, \perp} = \mathbf{\tilde k}_{i, \perp} - \frac{x_i}{x_V} \mathbf{k}_{V, \perp}$ and using the fact that 
\begin{equation}
\delta^{(2)} (\sum_{i=1}^3 \mathbf{k}_{i, \perp} + \mathbf{k}_{R, \perp}) = \delta^{(2)} (\sum_{i=1}^3 \mathbf{\tilde k}_{i, \perp}), 
\end{equation}
one factorizes the $d^2\tilde k_{R,\perp}$ and $d^2 k_{i,\perp}$ integrations resulting in a 
following expression for the transverse integrals of  Eq.(\ref{fqinp1}) :
\begin{align} 
&\int^{Q^2} [d^2\mathbf{k}_{\perp}] \exp \left[ -\frac{B_Vx_V}{4} \sum_{i=1}^3  
\frac{k_{i,\perp}^2}{x_i} - B_R k_{R,\perp}^2  \right]  \nonumber \\
&= \int^{Q^2} \prod_{i=1}^3 \frac{d^2 \mathbf{\tilde k}_{i, \perp}}{16 \pi^3} 16\pi^3 \delta^{(2)} 
\left(\sum_{i=1}^3 \mathbf{\tilde k}_{i, \perp} \right)\nonumber \\
&\times \exp \left[-\frac{B_V x_V}{4} \sum_{i=1}^3 \frac{\tilde k_{i, \perp}^2}{x_i}\right]     \int^{Q^2} 
\frac{d^2 \mathbf{k}_{R, \perp}}{16 \pi^3} e^{-B_R k_{R, \perp}^2}.
\label{term2}
\end{align}
Furthermore, we take the 
$d^2\mathbf{\tilde k}_{3,\perp}$ integral using the $\delta^2()$ function and 
evaluate the $d^2 \mathbf{k}_{R, \perp}$ integral analytically,  arriving at:
\begin{align}
& \frac{1}{(16\pi^3)^2} \int^{Q^2}  d^2 \mathbf{\tilde k}_{1, \perp} d^2 \mathbf{\tilde k}_{2, \perp}  \nonumber \\ 
&\times \exp{\left[-\frac{B_V x_V}{4} \left( \frac{\tilde k_{1, \perp}^2}{x_1} + 
\frac{\tilde k_{2, \perp}^2}{x_2} + \frac{(\mathbf{\tilde k}_{1,\perp} + \mathbf{\tilde k}_{2,\perp})^2}{x_3}\right) \right]} \nonumber \\ 
&\times  \frac{\pi}{16\pi^3} \int_0^{Q^2} dk_{R,\perp}^2 e^{-B_R k_{R, \perp}^2} \nonumber \\
&= \frac{1}{(16\pi^3)^2} \int^{Q^2}  d^2 \mathbf{\tilde k}_{1, \perp} d^2 \mathbf{\tilde k}_{2, \perp}  \nonumber \\  
&\times \exp{ \left[-\frac{B_V x_V}{4} \left( 
\frac{\tilde k_{1, \perp}^2}{x_1} + \frac{\tilde k_{2, \perp}^2}{x_2} + \frac{(\mathbf{\tilde k}_{1,\perp} + \mathbf{\tilde k}_{2,\perp})^2}{x_3}\right) \right]} \nonumber \\ 
&\times  \frac{\pi(1-e^{-B_R Q^2})}{16\pi^3 B_R}.
\label{term4}
\end{align}
Next we  introduce  CM and relative transverse momenta for ``1" and ``2" quarks:
\begin{eqnarray}
\mathbf{k}_{12}^{cm} &=& \mathbf{\tilde k}_{1, \perp} + \mathbf{\tilde k}_{2, \perp}, \quad \mathbf{\tilde k}_{12}^{rel} = \frac{x_2\mathbf{\tilde k}_{1, \perp} - x_1\mathbf{\tilde k}_{2, \perp}}{x_1 +x_2},
\end{eqnarray}
which allows  to decouple  the  momentum dependent term in Eq.(\ref{term4}) as  follows:
\begin{align}
&\frac{k_{1, \perp}^2}{x_1} + \frac{k_{2, \perp}^2}{x_2} + \frac{(\mathbf{k}_{1,\perp} + \mathbf{k}_{2,\perp})^2}{x_3} \nonumber \\   
&=  (k_{12}^{rel})^2 \left(\frac{1}{x_1} + \frac{1}{x_2} \right) + (k_{12}^{cm})^2 \left(  \frac{x_V}{x_3(x_1 + x_2)}  \right).
\end{align}
Using this  and the  relation  $d^2\mathbf{\tilde k}_{1,\perp} d^2\mathbf{\tilde k}_{2,\perp} = d^2\mathbf{k}_{12}^{cm} d^2\mathbf{k}_{12}^{rel}$ for Eq.(\ref{term2}) one obtains:
\begin{align} 
&\int^{Q^2} [d^2\mathbf{k}_{\perp}] \exp \left[ -\frac{B_Vx_V}{4} \sum_{i=1}^3 \frac{k_{i,\perp}^2}{x_i} - B_R k_{R,\perp}^2  \right]  \nonumber \\
&=  \int^{Q^2}  d^2\mathbf{k}_{12}^{cm} d^2\mathbf{k}_{12}^{rel}  \nonumber \\ 
&\times \exp \bigg[-\frac{B_V x_V}{4} \bigg(
(k_{12}^{rel})^2 \left(\frac{1}{x_1} + \frac{1}{x_2} \right) \nonumber \\
& + (k_{12}^{cm})^2 \frac{x_V}{x_3(x_1 + x_2)}
\bigg) \bigg] \nonumber \\
&\times  \frac{1}{(16\pi^3)^2} \frac{\pi(1-e^{-B_R Q^2})}{(16\pi^3) B_R}.
\label{term5}
\end{align}
Introducing $a_{rel} = \frac{B_V x_V }{4} \left( \frac{1}{x_1} + \frac{1}{x_2} \right)$ and $a_{cm} = \frac{B_V x_V}{4} \frac{x_V}{x_3(x_1 + x_2)} $, the remaining part of the integration in 
Eq.(\ref{term5}) can be evaluated analytically resulting in:
\begin{align}
&\int^{Q^2} [d^2\mathbf{k}_{\perp}] \exp \big[ -B_V \sum_{i=1}^3 x_V \frac{k_{i,\perp}^2}{x_i} - B_R k_{R,\perp}^2  \big] \nonumber \\
&=  \frac{1}{(16\pi^3)^2}  \frac{\pi(1-e^{-a_{cm}Q_{cm}^{max 2}})}{16 \pi^3 a_{cm}} \frac{\pi(1-e^{-a_{rel}Q_{rel}^{max 2}})}{16 \pi^3 a_{rel}}  \nonumber \\
&\times \frac{ \pi(1-e^{-B_R Q^2})}{16\pi^3 B_R} \nonumber \\
&=  \frac{x_1 x_2 x_3 }{(16\pi^3)^4 x_V^3 B_V^2 B_R} (1-e^{-a_{cm}Q_{cm}^{max 2}})(1-e^{-a_{rel}Q_{rel}^{max 2}}) \nonumber \\
&\times (1-e^{-B_R Q^2}),
\end{align}
where $Q_{cm}^{max}$ and $Q_{rel}^{max}$ denote the max values of $k_{12}^{cm}$ and $k_{12}^{rel}$, respectively.  Substituting the above expression into Eq.(\ref{fqinp1}) one obtains:
\begin{align}
& f_q(x_B, Q^2) = \mathcal{N} \int [dx]\delta(x_B-x_1)\nonumber  \\
&\times \exp \left[ -\frac{B_Vx_V}{4} \sum_{i=1}^3 \frac{ m_i^2}{x_i} - B_R M_N^2(x_R - \frac{M_R}{M_N})^2 \right]  \nonumber  \\
&\times   \frac{x_1 x_2^2 x_3^2 x_R}{x_V^3} \left(1-e^{-a_{cm}Q_{cm}^{max 2}}\right)\left(1-e^{-a_{rel}Q_{rel}^{max 2}}\right) \nonumber  \\
&\times\left(1-e^{-B_R Q^2}\right),
\end{align}
where $\mathcal{N}= {16\pi^3 A_V^2 A_R^2 m_N^3\over  B_R B_V^2} e^{{9\over 4} B_V m_q^2}$.
The two remaining Dirac delta functions  can be used to evaluate 
the $d x_1$ and $d x_R$ integrals. 

%\bibliographystyle{spphys}
%\bibliography{highx_mfmod}

\end{document}